\documentclass[seceq]{ptptex}

\usepackage{amsmath,amsfonts,latexsym,amssymb} 
\usepackage{verbatim,amsthm,graphicx} 
\usepackage{wrapft}

\def\A{{\cal A}}

\def\E{{\cal E}} 
\def\F{{\cal F}}

\def\K{{\cal K}}

\def\N{{\cal N}}

\def\linebreak{\hfill\break}

\def\bra<#1|{\langle #1\rvert} 
\def\ket|#1>{\lvert#1 \rangle} 
\def\braket<#1|#2>{\langle #1|#2 \rangle}

\def\pfrac#1#2{\left(\frac{#1}{#2}\right)} 
 
\def\tend{\rightarrow}

\def\therefore{\mbox{\setbox0=\hbox{X}\hbox{$\ldotp$}\raise0.7\ht0\hbox{$\ldotp$}\hbox{$\ldotp$}} \quad } 
\def\because{\mbox{\setbox0=\hbox{X}\raise0.7\ht0\hbox{$\ldotp$}\hbox{$\ldotp$}\raise0.7\ht0\hbox{$\ldotp$}}\kern0pt } 
 
\def\RF{{{\mathbb R}}}

\def\sl{{\rm sl}}

\def\Frac(#1/#2){\left(\frac{#1}{#2}\right)}

\def\Lie{\hbox{\rlap{$\cal L$}$-$}}

\def\Eq#1{\begin{equation} #1 \end{equation}} 
 
\def\Eqr#1{\begin{eqnarray} #1 \end{eqnarray}} 
 
\def\Eqrsub#1{\begin{subequations}
\Eqr{#1}\end{subequations}}
\def\Eqrsubl#1#2{\begin{subequations}\label{#1}
\Eqr{#2}\end{subequations}}
\def\Bitm{\begin{itemize}} 
\def\Eitm{\end{itemize}} 
\def\Blist#1#2{\begin{list}{#1}{\parsep=0pt \itemsep=0pt%
  \listparindent=0pt #2}} 
\def\Elist{\end{list}}

\def\THB{{\mathbb T}}
\def\VHB{{\mathbb V}}
\def\SHB{{\mathbb S}}

\title{Master equations for perturbations of generalised\\ 
static black holes with charge in higher dimensions}

\author{Hideo {\sc Kodama}$^{1,}$%
\footnote{E-mail: kodama@yukawa.kyoto-u.ac.jp} 
and Akihiro {\sc Ishibashi}$^{2,}$%
\footnote{E-mail: A.Ishibashi@damtp.cam.ac.uk}
}

\inst{$^1$Yukawa Institute for Theoretical Physics, Kyoto University \\
Kyoto  606-8502, Japan 
\\
$^2$D.A.M.T.P., Centre for Mathematical Sciences, University of Cambridge\\ 
Wilberforce Road, Cambridge CB3 0WA, United Kingdom} 

\abst{
We extend the formulation for perturbations of maximally symmetric 
black holes in higher dimensions developed by the present authors in 
a previous paper to a charged black hole background whose horizon is 
described by an Einstein manifold. For charged black holes, 
perturbations of electromagnetic fields are coupled to the vector 
and scalar modes of metric perturbations non-trivially. We show that 
by taking appropriate combinations of gauge-invariant variables for 
these perturbations, the perturbation equations for the 
Einstein-Maxwell system are reduced to two decoupled second-order 
wave equations describing the behaviour of the electromagnetic mode 
and the gravitational mode, for any value of the cosmological 
constant. These wave equations are transformed into 
Schr\"odinger-type ODEs through a Fourier transformation with 
respect to time. Using these equations, we investigate the stability 
of generalised black holes with charge. We also give explicit 
expressions for the source terms of these master equations with 
application to the emission problem of gravitational waves in mind. }

\begin{document}

\maketitle

\section{Introduction}

In recent years, motivated by developments in higher-dimensional 
unification theories, the behaviour of gravity in higher dimensions 
has become one of the major subjects in fundamental physics. In 
particular, the proposals of TeV gravity theories in the context of 
large extra dimensions%
\cite{Arkani-Hamed.N&Dimopoulos&Dvail1998,Antoniadis.I&&1998} and  
warped compactification%
\cite{Randall.L&Sundrum1999a,Randall.L&Sundrum1999b} have led to the 
speculation that higher-dimensional black holes might be produced in 
colliders\cite{Dimopoulos.S&Landsberg2001,%
Giddings.S&Scott2002,Giddings.S&Thomas2002} 
and in cosmic ray events\cite{Feng.J&Shapere2002,%
Anchordoqui.L&Goldberg2002,Emparan.R&Masip&Rattazzi2002,%
Ahn.E&&2003A}. Although 
fully nonlinear analysis of the classical and quantum dynamics of 
black holes will eventually be required to test this possibility by 
experiments, linear perturbation analysis is expected to give 
valuable information concerning some aspects of the problem, such as 
the stability of black holes, an estimation of gravitational 
emission during  black hole formation, and the determination of the 
greybody factor for quantum evaporation of black holes%
\cite{Giddings.S&Thomas2002,Anchordoqui.L&Goldberg2002,%
Cardoso.V&Dias&Lemos2003,Cavaglia.M2003A}. 
The linear perturbation theory of black holes can be used also in the 
quasi-normal mode 
analysis\cite{Cardoso.V&Dias&Lemos2003,Cardoso.V&Lemos2003A,%
Cardoso.V&Konoplya&Lemos2003A,%
Konoplya.R2003A,MaassenvandenBrink.A2003A,Suneeta.V2003A,%
Berti.E&Kokkotas2003A,Berti.E&&2003A,Cardoso.V&&2003A,%
Berti.E&Kokkotas&Papantonopoulos2003A} 
of the AdS/CFT issues and to obtain some insight into whether the 
uniqueness theorems of asymptotically flat regular black holes in 
four dimension (see Ref. \citen{Heusler.M1996B} for a review) and in 
higher dimensions\cite{Israel.W1967,Israel.W1968,
Hwang.S1998,Gibbons.G&Ida&Shiromizu2002a,%
Gibbons.G&Ida&Shiromizu2002b,Rogatko.M2002,Rogatko.M2003} 
can be extended to the asymptotically de Sitter and anti-de Sitter 
cases.

With the motivation provided by the developments described above, in 
a previous paper, 
Ref.~\citen{Kodama.H&Ishibashi2003A} (Paper I), we developed a 
formulation that reduces the linear perturbation analysis of 
generalised static black holes in higher-dimensional spacetimes with 
or without a cosmological constant to the study of a single 
second-order ODE of the Schr\"odinger-type for each type of 
perturbation. Here, a generalised black hole is considered to be a  
black hole whose horizon geometry is described by an Einstein 
metric. This includes a maximally symmetric black hole, i.e. a black 
hole whose horizon has a spatial section with constant curvature, 
such as a spherically symmetric black hole, as a special case. Then, 
in Ref.~\citen{Ishibashi.A&Kodama2003A} (Paper II), we studied the 
stability of such black holes using this formulation and proved the 
perturbative stability of asymptotically flat static black holes in 
higher dimensions as well as asymptotically de Sitter and anti-de 
Sitter static black holes in four dimensions. We also showed that 
the other types of maximally symmetric and static black holes might 
be unstable only for scalar-type perturbations.

One of the main purposes of the present paper is to extend the 
formulation given in Paper I to a generalised black hole with charge 
and analyse its stability. This extension is non-trivial, because 
perturbations of the metric and the electromagnetic field couple in 
the Einstein-Maxwell system. Hence, the main task is to show that 
the perturbation equations for the Einstein-Maxwell system can be 
transformed into two decoupled equations by an appropriate choice of 
the perturbation variables, as in the four-dimensional 
case\cite{Moncrief.V1974,Zerilli.F1974,Chandrasekhar.S1983B}. 
Since higher-dimensional unified theories 
based on string/M theories contain various $U(1)$ gauge fields, this 
extension is expected to be useful in studying generic black holes 
in unified theories.  

The other purpose of the present paper is to give explicit 
expressions for the source terms of the master equations. This 
information is necessary to apply the formulation to the estimation 
of graviton and photon emissions from mini-black holes formed by 
colliders. 

This paper is organized as follows. In the next section, we first 
make clear the basic assumptions regarding the unperturbed 
background, and then we give a general argument concerning the 
tensorial decomposition of perturbations. We also give basic 
formulas for the perturbation of electromagnetic fields that are 
used in later sections. Then, in the subsequent three sections, we 
derive decoupled master equations with a source for the 
Einstein-Maxwell system in a generalised static background with a 
static electric field for tensor-type,  vector-type and scalar-type 
perturbations. In \S6, using the formulations given in the previous 
sections, we analyse the stability of generalised static black holes 
with charge. Section 7 is devoted to summary and discussion.

\section{Background Spacetime and Perturbation}

In this section, we first explain the assumptions concerning the 
unperturbed background spacetimes considered in the present paper 
and present the basic formulas concerning them. Then, we give 
general arguments on the types of perturbations and the expansion of 
perturbations in harmonic tensors on the Einstein space by 
supplementing the argument of Gibbons and Hartnoll given in Ref. 
\citen{Gibbons.G&Hartnoll2002} with some fine points associated with 
scalar and vector 
perturbations. Finally, we give the basic perturbation equations of 
electromagnetic fields and formulas used in the subsequent sections.

\subsection{Unperturbed background}
\label{subsec:BG}

In the present paper, we assume that the background manifold has the 
structure ${\cal M}\approx {\cal N}^2 \times \K^n$ 
locally and its metric is given 
by 
\Eq{
ds^2= g_{ab}(y)dy^a dy^b+r^2(y)d\sigma_n^2.
\label{BG:metric}
}
Here, $d\sigma_n^2=\gamma_{ij}(z)dz^idz^j$ is the metric of the 
$n$-dimensional Einstein space $\K^n$ with
\Eq{
\hat R_{ij}=(n-1)K\gamma_{ij},
\label{EinsteinSpace}
}
where $\hat R_{ij}$ is the Ricci tensor of the metric $\gamma_{ij}$. 
When the metric \eqref{BG:metric} represents a black hole spacetime, 
the space $\K^n$ describes the structure of a spatial section of its 
horizon.  In the case in which $\K^n$ is a constant curvature space, 
$K$ denotes its sectional curvature, while in a generic case, $K$ is 
just a constant representing a local average of the sectional 
curvature.  Because an Einstein space of dimension smaller than four 
always has a constant curvature and is maximally symmetric, this 
difference arises only for $n\ge4$, or, equivalently, when the 
spacetime dimension $d=n+2$ is greater than or equal to 6. In the 
present paper, we assume that $\K^n$ is complete 
with respect to the Einstein metric, and we normalise $K$ so that 
$K=0,\pm1$.

The non-vanishing connection coefficients of the metric 
\eqref{BG:metric} are 
\Eq{
\Gamma^a_{bc}={}^2\!\Gamma^a_{bc}(y),\quad
\Gamma^i_{jk}=\hat\Gamma^i_{jk}(z),\quad
\Gamma^a_{ij}=-rD^a r \gamma_{ij}(z),\quad
\Gamma^i_{ja}=\frac{D_a r}{r}\delta^i_j,
\label{BG:Christoffel}
}
and the curvature tensors are
\Eqrsub{
&& R^a{}_{bcd}={}^2\! R^a{}_{bcd},\\
&& R^a{}_{ibj}=-\frac{D^aD_b r}{r}g_{ij},\\
&& R^i{}_{jkl}=\hat R^i{}_{jkl}
         -(Dr)^2(\delta^i_k\gamma_{jl}-\delta^i_l\gamma_{jk}),
}
where $\hat R^i{}_{jkl}$ is the curvature tensor of $\gamma_{ij}$. 
{}From this and \eqref{EinsteinSpace}, we obtain
\Eqrsubl{BG:Ricci}{
& R_{ab}&=\frac{1}{2}{}^2\! R g_{ab} -n\frac{D_aD_b r}{r},\\
& R_{ai}&=0,\\
& R_{ij}&= \left(- \frac{\Box r}{r}
        +(n-1)\frac{K-(Dr)^2}{r^2} \right) g_{ij},\\
& R &={}^2\! R
    -2n\frac{\Box r}{r}+n(n-1)\frac{K-(Dr)^2}{r^2},
}
where $D_a$, $\Box$ and ${}^2\! R$ are the covariant derivative, the 
D'Alermbertian and the scalar curvature for the metric 
$g_{ab}$, respectively. Thus, the Ricci tensor takes the same form 
as in the case in which $\K^n$ is maximally symmetric, as first 
pointed out by Birmingham\cite{Birmingham.D1999}. 

As the background source for the gravitational field, we consider an 
electromagnetic field whose field strength tensor $\F_{\mu\nu}$ has 
the structure
\Eq{
\F=\frac{1}{2}E_0 \epsilon_{ab}dy^a\wedge dy^b
   +\frac{1}{2}\F_{ij}dz^i\wedge dz^j.
\label{BG:EMfield}
}
Then, from the Maxwell equation $d\F=0$, we obtain
\Eq{
E_0=E_0(y),\quad
\F_{ij}=\F_{ij}(z),\quad 
\F_{[ij,k]}=0,
}
and from $\nabla_\nu \F^{\mu\nu}=0$, 
\Eqrsub{
&& 0=\nabla_\nu \F^{a\nu}=\frac{1}{r^n}\epsilon^{ab}D_b(r^nE_0),\\
&& 0=\nabla_\nu \F^{i\nu}=\hat D_j\F^{ij}. 
}
These equations imply that the electric field $E_0$ takes the 
Coulomb form, 
\Eq{
E_0=\frac{q}{r^n},
\label{BG:Efield}
}
and $\hat\F=\frac{1}{2}\F_{ij}(z)dz^i\wedge dz^j$ is a harmonic form 
on $\K^n$. In general, there may exist such a harmonic form that 
produces an energy-momentum tensor consistent with the structure of 
the Ricci tensors in \eqref{BG:Ricci}, if the second Betti number of 
$\K^n$ is not zero. The monopole-type magnetic field in 
four-dimensional spacetime provides such an example. However, since 
scalar and vector perturbations become coupled if such a background 
field exists, in the present paper we consider only the case 
$\F_{ij}=0$. 

With this assumption, the energy-momentum tensor for the 
electromagnetic field,
\Eq{
T_{\mu\nu}=\F_{\mu\alpha}\F_\nu{}^\alpha
           -\frac{1}{4}g_{\mu\nu}\F_{\alpha\beta}\F^{\alpha\beta},
}
is written 
\Eq{
T^a_b=-P \delta^a_b,\ T^i_j=P\delta^i_j;\quad
P=\frac{1}{2}E_0^2=\frac{q^2}{2r^{2n}},
\label{EMtensor:BG}
}
and the background Einstein equations, 
\Eq{
R_{\mu\nu}=\frac{2}{n}\Lambda g_{\mu\nu}
  +\kappa^2\left( T_{\mu\nu}-\frac{1}{n}T g_{\mu\nu} \right),
}
are reduced to
\Eqrsubl{BG:EinsteinEq}{
&& n\Box r=r\,{}^2\!R-2(n+1)\lambda r+\frac{2(n-1)^2Q^2}{r^{2n-1}},
\label{BG:EinsteinEq:1}\\
&& (n-1)\frac{K-(Dr)^2}{r^2}=\frac{\Box r}{r}+(n+1)\lambda
       +\frac{(n-1)Q^2}{r^{2n}},
\label{BG:EinsteinEq:2}\\
&& 2 D_aD_b r= \Box r g_{ab},
\label{BG:EinsteinEq:3}
}
where
\Eq{
\lambda:=\frac{2\Lambda}{n(n+1)},\quad
Q^2:=\frac{\kappa^2 q^2}{n(n-1)}.
}
{}From \eqref{BG:EinsteinEq:1}, \eqref{BG:EinsteinEq:3} and the 
identity
\Eq{
D_a\Box r=\Box D_ar -\frac{1}{2}{}^2\! R D_ar,
}
it follows that
\Eq{
D_a\left( r^{n+1}({}^2\! R-2\lambda)
  -\frac{2(n-1)(2n-1)Q^2}{r^{n-1}} \right)=0.
}
Hence, we obtain
\Eqrsub{
&& \frac{\Box r}{r}
 =-2\lambda+\frac{2(n-1)M}{r^{n+1}}-\frac{2(n-1)Q^2}{r^{2n}},\\
&& \frac{K-(Dr)^2}{r^2}
  =\lambda+\frac{2M}{r^{n+1}}-\frac{Q^2}{r^{2n}},\\
&& 
{}^2\!R=2\lambda+\frac{2n(n-1)M}{r^{n+1}}
     -\frac{2(n-1)(2n-1)Q^2}{r^{2n}}.
}

When $\nabla r\not=0$, these equations give the black hole type 
solution
\Eq{
ds^2=-f(r)dt^2+\frac{dr^2}{f(r)}+r^2d\sigma_n^2,
\label{metric:GSBH}
}
with
\Eq{
f(r)=K-\lambda r^2 -\frac{2M}{r^{n-1}}+\frac{Q^2}{r^{2n-2}}.
\label{f:RNBH}
}
To be precise, the spacetime described by this metric contains a 
regular black hole for $\lambda_c\le \lambda<0$ if $K=0$ or $K=-1$, 
and for $\lambda_{c1}\le \lambda < \lambda_{c2}$ and 
$Q^2/M^2<(n+1)^2/(4n)$ if $K=1$, where $\lambda_c,\lambda_{c1}$ and 
$\lambda_{c2}$ are functions of $Q,M$ and $n$. (For details, see 
Appendix~\ref{Appendix:D}.) 

Next, let us consider the case in which $r=a$ is constant. Here, we 
obtain the Nariai-type solution\cite{Nariai.H1950,Nariai.H1961} 
\Eq{
ds^2=-f(\rho)dt^2+\frac{d\rho^2}{f(\rho)}+a^2d\sigma_n^2,
\label{metric:Nariai}
}
where
\Eq{
f(\rho)=1-\sigma \rho^2;\quad
\sigma=(n+1)\lambda - \frac{(n-1)^2Q^2}{a^{2n}},
}
and the constant $a$ is determined as a solution to 
\Eq{
(n-1)\frac{K}{a^2}=(n+1)\lambda + \frac{(n-1)Q^2}{a^{2n}}.
}
%

\subsection{Tensorial decomposition of perturbations and the 
Einstein equations}\label{subsec:TensorialDecomposition}

In general, as tensors on $\K^n$, the metric perturbation variables 
$h_{\mu\nu}=\delta g_{\mu\nu}$ are classified into three groups of 
components, the scalar $h_{ab}$, the vector $h_{ai}$ and the tensor 
$h_{ij}$. Unfortunately, this grouping is not so useful, since 
components belonging to different groups are coupled through  
contraction with the metric tensor and the covariant derivatives in 
the Einstein equations. However, in the case that $\K^n$ is 
maximally symmetric, if we further decompose the vector and tensor as
\Eqrsubl{TensorialDecomposition}{
&& h_{ai}=\hat D_i h_a + h^{(1)}_{ai};\quad \hat D^ih^{(1)}_{ai}=0,\\
&& h_{ij}=h_L\gamma_{ij}+ h_{Tij};
\quad h_{T}{}^i_j=0,\\
&& h_{Tij}=\left(  \hat D_i\hat D_j
      -\frac{1}{n}\gamma_{ij}\hat\triangle\right)h^{(0)}_{T}
   +2\hat D_{(i}h^{(1)}_{T j)}+h^{(2)}_{T ij};\nonumber\\
&& \hat D^j h^{(1)}_{T j}=0,\quad
   \hat D^j h^{(2)}_{T ij}=0, 
}
the Einstein equations are decomposed into three groups, each of 
which contains only variables belonging to one of the three sets of 
variables $\{h_{ab}, h_a, h_L, h^{(0)}_T\}$, $\{h^{(1)}_{ai}, 
h^{(1)}_{Ti}\}$ and 
$\{h^{(2)}_{Tij}\}$\cite{Mukohyama.S2000a,Kodama.H&Ishibashi&Seto2000}.
 Variables belonging to each set are called the scalar-type, the 
vector-type and the tensor-type variables, respectively. This 
phenomenon arises because the metric tensor $\gamma_{ij}$ is the 
only non-trivial tensor in the maximally symmetric space, and as a 
consequence, the tensorial operations on $h_{\mu\nu}$ to construct 
the Einstein tensors preserve this 
decomposition\cite{Kodama.H&Sasaki1984}. Moreover, for the same 
reason, the covariant derivatives are always combined into the 
Laplacian in the Einstein equations after this decomposition. Thus, 
the harmonic expansion of the perturbation variables with respect to 
the Laplacian is useful. 

In the case in which $\K^n$ is of the Einstein type, the Laplacian 
preserves the transverse condition,
\Eq{
\hat D^j \hat D\cdot \hat D h^{(1)}_{Tj}=0,  
}
and \eqref{TensorialDecomposition} leads to the relations
\Eqrsubl{LoweringOperation}{
&& \hat D^i h_{ai}=\hat\triangle h_a,\\
&& \hat D^jh_{Tij}=\frac{n-1}{n}\hat D_i
        (\hat\triangle +nK)h^{(0)}_T
        +\left(\hat D\cdot\hat D +(n-1)K\right)h^{(1)}_{Ti},\\
&& \hat D^i\hat D^j h_{Tij}=\frac{n-1}{n}(\hat\triangle +nK)
        \hat\triangle h^{(0)}_T.
}
Hence, the tensorial decomposition \eqref{TensorialDecomposition} is 
still well-defined if these equations can be solved with respect to 
$h_a$, $h^{(0)}_T$ and $h^{(1)}_{Ti}$. We assume that this condition 
is satisfied in the present paper. 

The relations \eqref{LoweringOperation} also show that tensor 
operations on $h_{\mu\nu}$ that lower the rank as tensors on $\K^n$ 
preserve the tensorial decomposition into the scalar type and  
vector type, because the Weyl tensor $\hat C^i{}_{jkl}$ of $\K^n$ is 
of second order with respect to differentiation and does not take 
part in such operations. It is also clear that tensor 
operations on $h_{ab}$ and $h_{ai}$ that preserve or increase the 
rank have the same property. Furthermore, the covariant derivatives 
are always combined into the Laplacian in the tensor equations for 
the scalar-type and vector-type variables obtained through these 
operations. Therefore the difference between the perturbation 
equations in the Einstein case and the maximally symmetric case can 
arise only through the operations that produce the second-rank terms 
in $(\delta G_{ij})_T$ from $h_{ij}$. As shown by Gibbons and 
Hartnoll\cite{Gibbons.G&Hartnoll2002}, these terms are given by %
\Eq{
\hat \triangle_L h_{ij} :=-\hat D\cdot\hat D h_{ij}
     -2\hat R_{ikjl}h^{kl}+2(n-1)Kh_{ij}.
\label{LichnerowiczOp:def}
}
The Lichnerowicz operator $\hat\triangle_L$ defined by this equation 
preserves the transverse and trace-free property of $h_{ij}$ as 
pointed out in Ref.~\citen{Gibbons.G&Hartnoll2002}. Furthermore, we can easily check that 
the following relations hold:
\Eqrsub{
&& \hat\triangle_L \left(\hat D_i\hat D_j 
     -\frac{1}{n}\gamma_{ij}\hat\triangle \right)h^{(0)}_T
  =-\left(\hat D_i\hat D_j 
     -\frac{1}{n}\gamma_{ij}\hat\triangle \right)
     \hat\triangle h^{(0)}_T,\\
&& \hat\triangle_L \left(\hat D_{(i}h^{(1)}_{Tj)}\right)
  =-\hat D_{(i}\left( \hat D\cdot\hat D-(n-1)K \right)h^{(1)}_{Tj)}.
}
Hence, the Lichnerowicz operator also preserves the tensorial types. 
One can also see that 
only the Laplacian appears as a differential operator in the 
perturbed Einstein equations for the scalar-type and vector-type 
components after the tensorial decomposition. Thus, we can utilize 
the expansion in terms of scalar and vector harmonics for these 
types of perturbations, and the structure of the Einstein space 
affects the perturbation equations only through the spectrum of the 
Laplacian $\hat D\cdot\hat D$. Similarly, for tensor-type 
perturbations, if we expand the perturbation variable 
$h^{(2)}_{Tij}$ in the eigenfunctions of the Lichnerowicz operator, 
we obtain the perturbation equation for the Einstein case from that 
for the maximally symmetric case by replacing the eigenvalue $k_T^2$ 
for $-\hat D\cdot\hat D$ in the latter case by $\lambda_L-2nK$, 
where $\lambda_L$ is an eigenvalue of $\hat \triangle_L$.

\subsection{Perturbation of electromagnetic fields}
\label{subsec:EMFperturbation}

The Maxwell equations consist of two sets of equations for the 
electromagnetic field strength $\F=(1/2)\F_{\mu\nu}dx^\mu\wedge 
dx^\nu$, $d\F=0$ and $\nabla_\nu \F^{\mu\nu}=J^\mu$. If we regard 
$(\delta \F)_{\mu\nu}=\delta \F_{\mu\nu}$ as the basic perturbation 
variable, the perturbation of the first equation does not couple to 
the metric perturbation, and it is simply given by 
\Eq{
d\delta \F=0.
\label{MaxwellEq:Perturbation:dF}
}
This is simply the condition that $\delta\F$ is expressed in terms 
of the perturbation of the vector potential, $\delta \A$, as 
$\delta\F=d\delta\A$. Next, perturbation of the second equation 
gives two set of equations, 
\Eqrsubl{MaxwellEq:Perturbation}{
&& \frac{1}{r^n}D_b(r^n(\delta\F)^{ab}) 
          +\hat D_i(\delta\F)^{ai}
          +E_0\epsilon^{ab}\left( 
          \frac{1}{2}D_b (h^i_i-h^c_c)-\hat D_i h^i_b\right) =J^a,
\label{MaxwellEq:Perturbation:Ja}\\
&& \frac{1}{r^{n-2}}D_a\left[ r^{n-2}
      \left( (\delta\F)_i{}^a+E_0\epsilon^{ab}h_{ib} \right) \right]
       +\hat D_j(\delta\F)_i{}^j =J_i,
\label{MaxwellEq:Perturbation:Ji}
}
where $h_{\mu\nu}=\delta g_{\mu\nu}$, and the external current 
$J^\mu$ is treated as a first-order quantity. 

The perturbation variable $\delta\F$ transforms under an 
infinitesimal coordinate transformation $x^\mu \tend x^\mu 
+\bar\delta x^\mu$ as 
\Eq{
\bar\delta (\delta \F_{\mu\nu})
=-(\Lie_{\bar\delta x}\delta \F)_{\mu\nu}
=-\bar\delta x^\alpha\nabla_\alpha\F_{\mu\nu}
  +2\F_{\alpha[\mu}\nabla_{\nu]}\bar\delta x^\alpha.
}
To be explicit, writing $\bar\delta x^\mu$ as
\Eq{
\bar\delta y^a=T^a,\quad
\bar\delta z^i=L^i,
\label{GaugeTrf:coord}
}
we have
\Eqrsubl{GaugeTrf:EMF}{
&& \bar\delta(\delta\F_{ab})=-D_c(E_0 T^c)\epsilon_{ab},\\
&& \bar\delta(\delta\F_{ai})=-E_0 \epsilon_{ab} \hat D_i T^b,\\
&& \bar\delta(\delta\F_{ij})=0.
}

As in the case of the metric perturbation, the perturbation of the 
electromagnetic field, $\delta\F_{\mu\nu}$, can be decomposed into 
different tensorial types. The only difference is that no tensor 
perturbation exists for the electromagnetic field, because it can be 
described by a vector potential. After harmonic expansion, 
the Maxwell equations \eqref{MaxwellEq:Perturbation} are decomposed 
into decoupled gauge-invariant equations for each type in the 
background consisting of \eqref{metric:GSBH} and 
\eqref{metric:Nariai},  
because the Weyl tensor of $\K^n$ does not appear in the Maxwell 
equations. This gauge-invariant formulation is given in the 
subsequent sections for each type. Here, we only note that 
$\delta\F$ is gauge-invariant for a vector perturbation, since from 
\eqref{GaugeTrf:EMF} the gauge transformation of $\delta\F$ does not 
depend on $L^i$. In contrast, for a scalar perturbation, these 
perturbation variables must be combined with perturbation variables 
for the metric to construct a basis for gauge-invariant variables 
for the electromagnetic fields. 

Finally, we give expressions for the contribution of
electromagnetic field perturbations to the energy-momentum tensor: 
\Eqrsubl{EMFperturbation:EMtensor}{
& \delta T_{ab}= & \frac{E_0}{2}\left(\epsilon^{cd}\delta\F_{cd}
         +E_0 h^c_c  \right)g_{ab}
         -\frac{1}{2}E_0^2 h_{ab} ,\\
& \delta T^a_i= & E_0\epsilon^{ab} \delta\F_{ib},\\
& \delta T^i_j= &-\frac{E_0}{2}\left( \epsilon^{cd}\delta\F_{cd}
         +E_0 h^c_c  \right)\delta^i_j.        
}
%

\section{Tensor-type Perturbation}

Because an electromagnetic field perturbation does not have a 
tensor-type component, the electromagnetic fields enter the 
equations for a tensor perturbation only through their effect on the 
background geometry.

As explained in \S\ref{subsec:TensorialDecomposition}, tensor 
perturbations of the metric and the energy-momentum tensor can be 
expanded in terms of the eigentensors $\THB_{ij}$ of the 
Lichnerowicz operator $\hat\triangle_L$ defined by 
\eqref{LichnerowiczOp:def} as
\Eqrsub{
&& \delta g_{ab}=0,\ 
   \delta g_{ai}=0,\ 
   \delta g_{ij}=2r^2H_T(y) \THB_{ij},\\
&& \delta T_{ab}=0,\ 
   \delta T^a_i=0,\ 
   \delta T^i_j=\tau_T(y) \THB^i_j.
}
The expansion coefficients $H_T$ and $\tau_T$ themselves are 
gauge-invariant, and the Einstein equations for them are obtained 
from those in the maximally symmetric case considered in Ref. 
\citen{Kodama.H&Ishibashi&Seto2000} through the replacement $k_T^2 
\tend \lambda_L-2nK$. The result is expressed by the single equation 
\Eq{
\Box H_T +\frac{n}{r}Dr\cdot 
DH_T-\frac{\lambda_L-2(n-1)K}{r^2}H_T=-\kappa^2\tau_T,
\label{BasicEq:tensor}
}
where $\lambda_L$ is the eigenvalue of the Lichnerowicz operator,
\Eq{
\hat\triangle_L \THB_{ij}=\lambda_L\THB_{ij}.
}

For the Nariai-type background \eqref{metric:Nariai}, the 
above perturbation equation simplifies to
\Eq{
-\partial_t^2H_T +f\partial_\rho (f\partial_\rho H_T) 
  -\frac{\lambda_L-2(n-1)K}{a^2}f H_T=-\kappa^2 f\tau_T.
}
For the black hole background \eqref{metric:GSBH}, if we introduce 
the master variable $\Phi$ by
\Eq{
\Phi=r^{n/2}H_T,
}
\eqref{BasicEq:tensor} can be put into the canonical form
\Eq{
 \Box\Phi- \frac{V_T}{f}\Phi=-\kappa^2 r^{n/2} \tau_T, 
\label{MasterEq:Tensor}
}
where
\Eq{
V_T= \frac{f}{r^2}\left[ 
                        \lambda_L-2(n-1)K
                        +\frac{nrf'}{2}+\frac{n(n-2)f}{4}
                  \right]. 
\label{VT:GSBH}
}
In particular, for $f(r)$ given by \eqref{f:RNBH}, $V_T$ is expressed as
\Eq{
V_T=\frac{f}{r^2}\left[\lambda_L +\frac{n^2-10n+8}{4}K 
    -\frac{n(n+2)}{4}\lambda r^2+\frac{n^2M}{2r^{n-1}}
    -\frac{n(3n-2)Q^2}{4r^{2n-2}}
 \right].
\label{VT:RNBH}
}

Note that \eqref{MasterEq:Tensor} with \eqref{VT:GSBH} describes the 
behaviour of a tensor perturbation if the background metric has the 
form \eqref{metric:GSBH}. Hence, it may apply to a system more 
general than the Einstein-Maxwell system considered in the present 
paper. 

\section{Vector-type Perturbation}

We expand vector perturbations in vector harmonics $\VHB_i$ 
satisfying
\Eq{
(\hat D\cdot\hat D +k_V^2)\VHB_i=0;\quad \hat{D}^i \VHB_i=0}
and the symmetric trace-free tensors $\VHB_{ij}$ defined by
\Eq{
\VHB_{ij}=-\frac{1}{k_V}\hat D_{(i}\VHB_{j)}.
\label{VectorHarmonics:SymmetricTensor:def}
}
Note that this tensor is an eigentensor of the Lichnerowicz operator,
\Eq{
\hat\triangle_L \VHB_{ij}=\left[ k_V^2+(n-1)K \right]\VHB_{ij},
}
but it is not an eigentensor of the Laplacian in general. 

In this paper, we assume that the Laplacian $-\hat D\cdot\hat D$ is 
extended to a non-negative self-adjoint operator in the $L_2$-space 
of divergence-free vector fields on $\K^n$, in order to guarantee 
the completeness of the vector harmonics. Because $-\hat D\cdot\hat 
D$ is symmetric and non-negative in the space consisting of smooth 
divergence-free vector fields with compact support, it always 
possesses a  Friedrichs extension that has the desired 
property\cite{Akhiezer.N&Glazman1966B}. With this assumption, 
$k_V^2$ is non-negative.  

One subtlety that arises in this harmonic expansion concerns the 
zero modes of the Laplacian. If $\K^n$ is closed, from the 
integration of the identity $\hat D^i(V^j\hat D_i V_j)-\hat D^i V^j 
\hat D_i V_j=V^j\hat D\cdot\hat D V_j$, it follows that $\hat D_i 
\VHB_j=0$ for $k_V^2=0$. Hence, we cannot construct a harmonic 
tensor from such a vector harmonic. We obtain the same result even 
in the case in which $\K^n$ is open if we require that $\VHB^j\hat 
D_i \VHB_j$ fall off sufficiently rapidly at infinity. In the 
present paper, we assume that this fall-off condition is satisfied.  
From the identity $\hat D^j\hat D_i V_j=\hat D_i\hat 
D^jV_j+(n-1)K\hat V_i$, such a zero mode exists only in the case 
$K=0$. We can further show that vector fields satisfying $\hat D_i 
V_j=0$ exist if and only if $\K^n$ is a product of a locally flat 
space and an Einstein manifold with vanishing Ricci tensor.

More generally, $\VHB_{ij}$ vanishes if $\VHB_i$ is a Killing 
vector. In this case, from the relation
\Eq{
2k_V \hat D_j \VHB^j_i= \left[ k_V^2 -(n-1)K\right] \VHB_i,
}
$k_V^2$ takes the special value $k_V^2=(n-1)K$. Because $k_V^2\ge0$, 
this occurs only for $K=0$ or $K=1$. In the case $K=0$, this mode 
corresponds to the zero mode discussed above.  In the case $K=1$, 
since we are assuming that $\K^n$ is complete, $\K^n$ is compact and 
closed, as known from Myers' theorem\cite{Myers.S1941}, and we can 
show the converse, i.e. that if $k_V^2=(n-1)K$, then $\VHB_{ij}$ 
vanishes, by integrating the identity $0= {{\VHB}^i}^* \hat{D}^j 
\VHB_{ij}=\hat{D}^j({{\VHB}^i}^*\VHB_{ij})+k_V 
{{\VHB}^{ij}}^*\VHB_{ij}$ over $\K^n$. 
Furthermore, using the same identity for ${{\VHB}^i}^* \hat{D}^j 
\VHB_{ij}$, we can show that there is no 
eigenvalue in the range $0<k_V^2<n-1$\cite{Gibbons.G&Hartnoll2002}.

\subsection{The Einstein equations with a source}

In terms of vector harmonics, perturbations of the metric and the 
energy-momentum tensor can be expanded as
\Eqrsubl{VectorPerturbation:metric}{
&& \delta g_{ab}=0,\ 
   \delta g_{ai}=rf_a \VHB_i,\ 
   \delta g_{ij}=2r^2H_T \VHB_{ij},\\
&& \delta T_{ab}=0,\ 
   \delta T^a_i=r\tau^a \VHB_i,\ 
   \delta T^i_j=\tau_T \VHB^i_j.
}

For $m_V\equiv k^2-(n-1)K\not=0$, the matter variables $\tau_a$ and 
$\tau_T$ are themselves gauge-invariant, and the combination%
\Eq{
F_a=f_a + rD_a\left( \frac{H_T}{k_V} \right)
} 
can be adopted as a basis for gauge-invariant variables for the 
metric perturbation. The perturbation of the Einstein equations 
reduces to 
\Eqrsubl{VectorPerturbation:BasicEq:metric}{
&&D_a\left( r^{n+1}F^{(1)} \right)
  -m_V r^{n-1}\epsilon_{ab}F^b
  =-2\kappa^2 r^{n+1}\epsilon_{ab}\tau^b,
\label{BasicEq:metric:vector1}\\
&& k_V D_a(r^{n-1}F^a)=-\kappa^2r^n\tau_T,
\label{BasicEq:metric:vector2}
}
where $\epsilon_{ab}$ is the two-dimensional Levi-Civita tensor for 
$g_{ab}$, and
\Eq{
F^{(1)}=\epsilon^{ab}r D_a\pfrac{F_b}{r}
       =\epsilon^{ab}r D_a\pfrac{f_b}{r}.
\label{F^(1):def}
}
This should be supplemented by the perturbation of the 
energy-momentum conservation law
\Eq{
D_a(r^{n+1}\tau^a)+\frac{m_V}{2k_V}r^n\tau_T=0.
\label{EMconservation:vector}
}

Now, note that, for $m_V=0$, the perturbation variables $H_T$ and 
$\tau_T$ do not exist. The matter variable $\tau_a$ is still 
gauge-invariant, but concerning the metric variables, only the 
combination $F^{(1)}$ defined in terms of $f_a$ in \eqref{F^(1):def} 
is gauge invariant. In this case, the Einstein equations are reduced 
to the single equation \eqref{BasicEq:metric:vector1}, and the 
energy-momentum conservation law is given by 
\eqref{EMconservation:vector} without the $\tau_T$ term.

Now, we show that these gauge-invariant perturbation equations can 
be reduced to a single wave equation with a source in the 
two-dimensional spacetime $\N^2$. We first treat the case 
$m_V\not=0$. In this case, by eliminating $\tau_T$ in 
\eqref{BasicEq:metric:vector2} with the help of 
\eqref{EMconservation:vector}, we obtain 
\Eq{
D_a(r^{n-1}F^a)=\frac{2\kappa^2}{m_V} D_a(r^{n+1}\tau^a).
}
{}From this it follows that $F^a$ can be written in terms of a 
variable $\tilde \Omega$ as
\Eq{
r^{n-1}F^a=\epsilon^{ab}D_b\tilde\Omega
    +\frac{2\kappa^2}{m_V}r^{n+1}\tau^a.
\label{FaByOmega}
}
Inserting this expression into \eqref{BasicEq:metric:vector1}, we 
obtain the master equation
\Eq{
r^nD_a\left( \frac{1}{r^n}D^a\tilde\Omega \right)
  -\frac{m_V}{r^2}\tilde\Omega 
  =-\frac{2\kappa^2}{m_V} r^n\epsilon^{ab}D_a(r\tau_b).
\label{MasterEq:vector:metric}
}

Next, we consider the special modes with $m_V=0$. For these modes, 
from \eqref{EMconservation:vector} with $\tau_T=0$, it follows that 
$\tau_a$ can be expressed in terms of a function $\tau^{(1)}$ as
\Eq{
r^{n+1} \tau_a=\epsilon_{ab}D^b\tau^{(1)}.
\label{tau1:def}
}
Inserting this expression into \eqref{BasicEq:metric:vector1} with 
$\epsilon^{cd}D_c(F_d/r)$ replaced by $F^{(1)}/r$, we obtain
\Eq{
D_a(r^{n+1}F^{(1)})=-2\kappa^2D_a\tau^{(1)}.
} 
Taking account of the freedom of adding a constant in the definition 
of $\tau^{(1)}$, the general solution can be written
\Eq{
F^{(1)}=-\frac{2\kappa^2 \tau^{(1)}}{r^{n+1}}.
\label{BasicEq:vector:metric:exceptional}
}
Hence, there exists no dynamical freedom in these special modes. In 
particular, in the source-free case in which $\tau^{(1)}$ is a 
constant and $K=1$, this solution corresponds to adding a small 
rotation to the background solution.

\subsection{Einstein-Maxwell system}

As shown in \S\ref{subsec:EMFperturbation}, all components of 
$\delta\F_{\mu\nu}$ are invariant under a coordinate gauge 
transformation for a vector perturbation. In order to find an 
independent gauge-invariant variable, we expand $\delta\F_{ai}$ in 
vector harmonics as $\delta\F_{ai}=\A_a\VHB_i$. Then, since 
$\delta\F_{ab}=0$ for a vector perturbation, the 
$(a,b,i)$-component of the Maxwell equation 
\eqref{MaxwellEq:Perturbation:dF} is written
\Eq{
D_{[a}\A_{b]}=0.
}
Hence, $\A_a$ can be expressed in terms of a function $\A$ as 
$\A_a=D_a\A$. Then, the $(a,i,j)$-component of 
\eqref{MaxwellEq:Perturbation:dF} is written
\Eq{
D_a\delta \F_{ij}=D_a\A (\partial_i\VHB_j-\partial_j\VHB_i).
}
This implies that $\delta\F_{ij}$ can be expressed as
\Eq{
\delta\F_{ij}=C_{ij}+\A(\partial_i\VHB_j-\partial_j\VHB_i),
}
where $C_{ij}$ is an antisymmetric tensor on $\K^n$ that does not 
depend on the $y$-coordinates. Finally, from the $(i,j,k)$-component 
of \eqref{MaxwellEq:Perturbation:dF}, it follows  that 
$C_{ij}dz^i\wedge dz^j$ is a closed form on $\K^n$. Hence, $C_{ij}$ 
can be expressed in terms of a divergence-free vector field $W_i$ as 
$C_{ij}=\partial_i W_j-\partial_j W_i$. With the expansion in vector 
harmonics, we can assume that $W_i$ is a constant multiple of 
$\VHB_i$, without loss of generality. Hence, this term can be 
absorbed into $\A$ by redefining it through the addition of a 
constant. Thus, we find a vector perturbation of the electromagnetic 
field can be expressed in terms of the single gauge-invariant 
variable $\A$ as%
\Eq{
\delta \F_{ab}=0,\ 
\delta \F_{ai}=D_a\A\VHB_i,\ 
\delta \F_{ij}=\A \left(\hat D_i\VHB_j-\hat D_j\VHB_i\right).
}

Next, we express the remaining Maxwell equations in terms of this 
gauge-invariant variable. For a vector perturbation, 
only Eq.\eqref{MaxwellEq:Perturbation:Ji} is non-trivial.  If we 
expand the current $J_i$ as 
\Eq{
J_i=J\VHB_i,
}
it can be written
\Eq{
 -\frac{1}{r^{n-2}}D_a\left(r^{n-2}D^a\A
       -r^{n-1}E_0\epsilon^{ab}f_b\right)\VHB_i 
 +\A \hat{D}^j\left(\hat D_i\VHB_j-\hat D_j\VHB_i\right) =J\VHB_i. 
}
Hence, from the identity
\Eq{
 \hat{D}^j \left(\hat D_i\VHB_j-\hat D_j\VHB_i\right)
  =\frac{k_V^2+(n-1)K}{r^2}\VHB_i,
}
the gauge-invariant form for the Maxwell equation is given by
\Eq{
 \frac{1}{r^{n-2}}D_a(r^{n-2}D^a \A)
    -\frac{k_V^2+(n-1)K}{r^2}\A  
   =-J +r E_0 F^{(1)}.
\label{RN:vector:Master:EM0}
}

In order to complete the formulation of the  basic perturbation 
equations, we must separate the contribution of the electromagnetic 
field to the source term in the Einstein equation 
\eqref{MasterEq:vector:metric}. Because for a vector perturbation, 
\eqref{EMFperturbation:EMtensor} is expressed as
\Eq{
\delta T^a_b=0,\quad
\delta T^a_i=-E_0 \epsilon^{ab}D_b\A\VHB_i,\quad
\delta T^i_j=0,
}
the contributions of the electromagnetic field to $\tau_a$ and 
$\tau_T$ are given by
\Eq{
\tau_{a}^{\rm{EM}}=-\frac{E_0}{r}\epsilon_{ab}D^b\A,\quad
\tau_{T}^{\rm{EM}}=0.
\label{tau:vector:EM}
}
Hence, the Einstein equations for the Einstein-Maxwell system can be 
obtained by replacing $\tau_a$ in \eqref{MasterEq:vector:metric} by
\Eq{
\tau_a=\tau_{a}^{\rm{EM}}+ \bar\tau_a,
}
where the second term represents the contribution from matter other 
than the electromagnetic field. 

In order to rewrite the basic equations obtained to this point in  
simpler forms, we treat the generic modes and the exceptional modes 
separately.

\subsubsection{Generic modes}

For modes with $m_V\equiv k_V^2-(n-1)K\not=0$, the above simple 
replacement yields a wave equation for $\tilde\Omega$ with $\Box\A$ 
in the source term. This second derivative term can be eliminated if 
we extract the contribution of the electromagnetic field to $\tilde 
\Omega$ as
\Eq{
\Omega=\tilde\Omega -\frac{2\kappa^2 q}{m_V}\A.
}
The result is
\Eq{
 r^nD_a\left( \frac{1}{r^n}D^a\Omega \right)
  -\frac{m_V}{r^2}\Omega 
  =\frac{2\kappa^2 q}{r^2}\A
   -\frac{2\kappa^2}{m_V} r^n\epsilon^{ab}D_a(r\bar\tau_b).
\label{RN:vector:Master:metric}
}
Equations \eqref{FaByOmega} and \eqref{EMconservation:vector} are 
replaced by
\Eqrsub{
&& r^{n-1}F^a=\epsilon^{ab}D_b \Omega
    +\frac{2\kappa^2}{m_V}r^{n+1}\bar\tau^a,\\
\label{FaByOmega:EM}
&& D_a(r^{n+1}\bar \tau^a)+\frac{m_V}{2k_V}r^n\tau_T=0.
\label{EMconservation:vector:EM}
}
Finally, inserting this expression for $F^a$ into 
\eqref{RN:vector:Master:EM0} and using 
\eqref{RN:vector:Master:metric}, we obtain
\Eq{
\frac{1}{r^{n-2}}D_a(r^{n-2}D^a\A)-\frac{1}{r^2}
 \left(k_V^2+(n-1)K + \frac{2n(n-1)Q^2}{r^{2n-2}}\right)\A 
=\frac{qm_V}{r^{2n}}\Omega -J.
\label{RN:vector:Master:EM}
}
Thus, the coupled system of equations consisting of 
\eqref{RN:vector:Master:metric} and \eqref{RN:vector:Master:EM} 
provides the basic gauge-invariant equations with source for a 
vector perturbation with $k_V^2\not=(n-1)K$. 

\subsubsection{Exceptional modes}
 
For exceptional modes with $k_V^2=(n-1)K$, from the definition 
\eqref{tau1:def} of $\tau^{(1)}$ and \eqref{tau:vector:EM}, 
$\tau^{(1)}$ can be expressed as
\Eq{
\tau^{(1)}=-q\A +\bar\tau^{(1)}.
}
Hence, \eqref{BasicEq:vector:metric:exceptional} can be rewritten as
\Eq{
F^{(1)}=\frac{2\kappa^2(q\A-\bar\tau^{(1)})}{r^{n+1}}.
}
Inserting this into \eqref{RN:vector:Master:EM0}, we obtain
\Eq{
\frac{1}{r^{n-2}}D_a(r^{n-2}D^a\A)-\frac{1}{r^2}
 \left(2(n-1)K + \frac{2n(n-1)Q^2}{r^{2n-2}}\right)\A 
= -J -\frac{2\kappa^2 q}{r^{2n}}\bar\tau^{(1)}.
\label{RN:vector:Master:EM:exceptional}
}
Therefore, only the electromagnetic perturbation is dynamical.

\subsection{Decoupled master equations}

The basic equations for the generic modes are coupled differential 
equations and are not useful. Fortunately, they can be transformed 
into a set of two decoupled equations by simply introducing master 
variables written as linear combinations of the original 
gauge-invariant variables. Such combinations can be found through 
simple algebraic manipulations. 

\subsubsection{Black hole background} 

For the black hole background \eqref{metric:GSBH}, appropriate combinations are given by
\Eq{
\Phi_\pm =a_\pm r^{-n/2}\Omega + b_\pm r^{n/2-1}\A,
}
with 
\Eqrsub{
&& (a_+,b_+)= 
  \left( \frac{Q m_V}{(n^2-1)M+\Delta},\frac{Q}{q} \right),\\
&& (a_-,b_-)= 
   \left(1,\frac{-2n(n-1)Q^2}{q[(n^2-1)M+\Delta]} \right),
}
where $\Delta$ is a positive constant satisfying 
\Eq{
\Delta^2= (n^2-1)^2M^2+2n(n-1)m_VQ^2.
}
When expressed in terms of these master variables, 
\eqref{RN:vector:Master:metric} and \eqref{RN:vector:Master:EM} are 
transformed into the two decoupled wave equations
\Eq{ 
  \Box \Phi_\pm - \frac{V_{V\pm}}{f}\Phi_\pm 
  =S_{V\pm}.
}
%
Here, 
\Eq{
V_{V\pm}=\frac{f}{r^2}\left[k_V^2 +\frac{(n^2-2n+4)K}{4}
   -\frac{n(n-2)}{4}\lambda r^2+\frac{n(5n-2)Q^2}{4r^{2n-2}} 
   +\frac{\mu_\pm}{r^{n-1}}\right],
\label{Vpm:Vector}
}
where
\Eq{
\mu_\pm=-\frac{n^2+2}{2}M \pm \Delta,
}
and 
\Eq{
S_{V\pm}=-a_\pm \frac{2\kappa^2 r^{n/2}f}{m_V}
      \epsilon^{ab}D_a(r\bar\tau_b)
    -b_\pm r^{n/2-1}f J.
}
For $n=2, K=1$ and $\lambda=0$, the variables $\Phi_+$ and $\Phi_-$ 
are proportional to the variables for the axial modes, $Z^{(-)}_1$ 
and $Z^{(-)}_2$ given in Ref. \citen{Chandrasekhar.S1983B}, and 
$V_{V+}$ and $V_{V-}$ coincide with the corresponding potentials, 
$V^{(-)}_1$ and $V^{(-)}_2$, respectively.

Here, note that in the limit $Q\tend0$, $\Phi_+$ becomes 
proportional to $\A$ and $\Phi_-$ to $\Omega$. Hence, $\Phi_+$ and 
$\Phi_-$ represent the electromagnetic mode and the gravitational 
mode, respectively. In particular, in the limit $Q\tend0$, the 
equation for $\Phi_-$ coincides with the master equation for a 
vector perturbation on a neutral black hole background derived in 
Paper I.

\subsubsection{Nariai-type background}

For the Nariai-type background \eqref{metric:Nariai}, the combinations
\Eq{
\Phi_\pm = a_\pm \Omega + b_\pm \A,
}
with
\Eqrsub{
&& (a_+,b_+)=\left( \frac{m_V Q}{a^{2n-2}\left[   2(n-1)K-a^2\sigma+a^2\Delta_N\right]} ,\frac{Q}{q}\right),\\
&& (a_-,b_-)=\left( 1, 
 -\frac{2n(n-1)Q^2}{q\left[ 2(n-1)K-a^2\sigma+a^2\Delta_N \right]} 
 \right),
}
give the decoupled equations
\Eqr{
&&-\partial_t^2\Phi_\pm +f\partial_\rho(f\partial_\rho\Phi_\pm)
-f\left( \frac{k_V^2+(n-1)K}{a^2}-\sigma \pm \Delta_N \right)\Phi_\pm
\notag\\
&& \quad = -\frac{2\kappa^2}{m_V}a_\pm f a^{n+1}\epsilon^{bc}D_b\bar\tau_c
   -b_\pm f J,
}
where
\Eq{
\Delta_N=\left[ \sigma^2
  +\frac{2n(n-1)Q^2}{a^{2n+2}}\left\{ k_V^2+(n-1)K \right\} 
  \right]^{1/2}.
}
%

\section{Scalar-type Perturbation}
\label{sec:ScalarPerturbation}

We expand scalar perturbations in scalar harmonics satisfying
\Eq{
\left( \hat\triangle +k^2 \right)\SHB=0.
}
As in the case of vector harmonics, we assume that $-\hat\triangle$ 
is extended to a non-negative self-adjoint operator in the 
$L^2$-space of functions on $\K^n$. Hence, $k^2\ge0$. In the present 
case, such an extension is unique, since we are assuming that $\K^n$ 
is complete\cite{Craioveanu.M&Puta&Rassias2001B}, and it is given by 
the Friedrichs self-adjoint extension of the symmetric and 
non-negative operator $-\hat\triangle$ on $C_0^\infty(\K^n)$. 
If $\K^n$ is closed, the spectrum of $\hat\triangle$ is 
completely discrete, each eigenvalue has a finite multiplicity, and 
the lowest eigenvalue is $k^2$=0, whose 
eigenfunction is a constant. A perturbation corresponding to such a 
constant mode represents a variation of the parameters $\lambda, M$, 
and $Q$ of the background, as seen from the argument in 
\S\ref{subsec:BG}. For this reason, we do not consider the modes 
with $k^2=0$, although there may be a 
non-trivial eigenfunction with $k^2=0$ in the cases in which $\K^n$ 
is open. Note that when $k^2=0$ is contained in the full spectrum 
but does not belong to the point spectrum, as in the case 
$\K^n=\RF^n$, it can be ignored without loss of generality.  

For modes with $k^2>0$, we can use the vector fields and the 
symmetric trace-free tensor fields defined by 
\Eqrsub{
&& \SHB_i=-\frac{1}{k} \hat D_i\SHB,\\
&& \SHB_{ij}=\frac{1}{k^2}\hat D_i\hat D_j\SHB
             +\frac{1}{n}\gamma_{ij}\SHB;\ \SHB^i_i=0
}
to expand vector and symmetric trace-free tensor fields, 
respectively. Note that $\SHB_i$ is also an eigenmode of the 
operator $\hat D\cdot\hat D$, i.e., 
\Eq{
[\hat D\cdot\hat D+k^2-(n-1)K]\SHB_i=0,
}
while $\SHB_{ij}$ is an eigenmode of the Lichnerowicz operator:
\Eq{
(\hat\triangle_L -k^2)\SHB_{ij}=0.
}

In the case of scalar harmonics, the modes with $k^2=nK$ are 
exceptional. Given our assumption, these modes exist only for $K=1$. 
Because $\K^n$ is compact and closed in this case, from the identity
\Eq{
\hat D_j \SHB^j_i=\frac{n-1}{n}\frac{k^2-nK}{k}\SHB_i,
}
we have $\hat D_j\SHB^j_i=0$. From this, it follows that $\int d^nz 
\sqrt{\gamma}{\SHB_{ij}}^*\SHB^{ij}=0$. Hence, $\SHB_{ij}$ vanishes 
identically.  

\subsection{Metric perturbations}

In terms of scalar harmonics, perturbations of the metric and the 
energy-momentum tensor are expanded as 
\Eqrsub{
&& \delta g_{ab}=f_{ab}\SHB,\ 
   \delta g_{ai}=rf_a \SHB_i,\ 
   \delta g_{ij}=2r^2(H_L\gamma_{ij}\SHB +H_T\SHB_{ij}),\\
&& \delta T_{ab}=\tau_{ab}\SHB,\ 
   \delta T^a_i=r\tau^a\SHB_i,\ 
   \delta T^i_j=\delta P \delta^i_j\SHB +\tau_T\SHB^i_j.
}
Following Ref. \citen{Kodama.H&Ishibashi&Seto2000}, we adopt the 
following combinations of these expansion coefficients as the basic 
gauge-invariant variables for perturbations of the metric and the 
energy-momentum tensor: 
\Eqrsubl{GaugeInvVar:scalar:metric}{
&& F=H_L+\frac{1}{n}H_T+\frac{1}{r}D^ar X_a,\\
&& F_{ab}=f_{ab}+D_aX_b+D_bX_a,\\
&& \Sigma_{ab}=\tau_{ab}-P(D_aX_b+D_bX_a)-X^cD_c P g_{ab},
\label{Sigma_ab:def}\\
&& \Sigma_a=\tau_a +\frac{2k}{r}PX_a,
\label{Sigma_a:def}\\
&& \Sigma_L = \delta {P} +X^aD_a{P}.
\label{Sigma_L:def}
}
Here, 
\Eq{
X_a=\frac{r}{k}\left(f_a+\frac{r}{k}D_a H_T\right),
\label{Xa:def}
}
and we have used the background value for $T^a_b$ given in \eqref{EMtensor:BG}. 

Now, note that for the exceptional modes with $k^2=n$ for $K=1$, 
$H_T$ and $\tau_T$ are not defined, because a second-rank symmetric 
tensor cannot be constructed from $\SHB$, as mentioned above. For 
such a mode, we define $F, F_{ab}, \Sigma_{ab},\Sigma_a$ and 
$\Sigma_L$ by setting $H_T=0$ in the above definitions. These 
quantities defined in this way are, however, gauge dependent. These 
exceptional modes are treated in Appendix \ref{Appendix:C}.

\subsection{Maxwell equations}

Because $X_a$ defined in \eqref{Xa:def} transforms under 
\eqref{GaugeTrf:coord} as
\Eq{
\bar\delta X_a=T_a,
}
from \eqref{GaugeTrf:EMF} it follows that the following combinations 
$\E$ and $\E_a$ can be used as basic gauge-invariants for 
perturbations of the electromagnetic field:
\Eqrsubl{GaugeInvVar:scalar:EMF}{
&& \delta \F_{ab}+D_c(E_0X^c)\epsilon_{ab}\SHB=\E \epsilon_{ab}\SHB,\\
&& \delta\F_{ai}-kE_0\epsilon_{ab}X^b\SHB_i=r\epsilon_{ab}\E^b\SHB_i.
}
Here, note that $\delta\F_{ij}=0$ for a scalar perturbation, since 
$\delta\F_{ij}$ can be written as $\delta\F_{ij}=\hat 
D_i\delta\A_j-\hat D_j\delta\A_i$, from the Maxwell equations, and 
$\delta\A_i$ is just the gradient of a scalar perturbation. 

By expanding the current $J_i$ as
\Eq{
J_i=rJ \SHB_i,
}
the Maxwell equations \eqref{MaxwellEq:Perturbation} can be written
\Eqrsub{
&& \frac{1}{r^n}D_a(r^n\E)+\frac{k}{r}\E_a
  -\frac{E_0}{2}D_a(F^c_c-2nF)=\epsilon_{ab}J^b,
\label{BasicEq:EM:Scalar:Maxwell1}\\
&& \epsilon^{ab}D_a(r^{n-1}\E_b)=-r^{n-1}J,
\label{BasicEq:EM:Scalar:Maxwell2}
}
while \eqref{MaxwellEq:Perturbation:dF} is expressed as
\Eq{
\E=-\frac{1}{k}D_c(r\E^c).
\label{BasicEq:EM:Scalar:EbyEa}
}
Note that \eqref{BasicEq:EM:Scalar:Maxwell1} and 
\eqref{BasicEq:EM:Scalar:Maxwell2} give the current conservation law
\Eq{
D_c(r^n J^c)=-kr^{n-1}J.
\label{BasicEq:EM:Scalar:J}
}

We can reduce these equations to a single wave equation for a single 
master variable. First, from \eqref{BasicEq:EM:Scalar:Maxwell2} and 
\eqref{BasicEq:EM:Scalar:J}, we have
\Eq{
\epsilon^{ab}D_a(r^{n-1}\E_b)=\frac{1}{k}D_c(r^n J^c).
}
Therefore, if we define $\tilde J_a$ by
\Eq{
J^a=\frac{k^2}{r^n}\epsilon^{ab}\tilde J_b,
}
$\E_a$ can be expressed in terms of a function $\A$ as
\Eq{
\E_a =\frac{k}{r^{n-1}}\left(D_a \A + \tilde J_a\right).
\label{BasicEq:EM:Scalar:EabyA}
}
Then, the insertion of this into \eqref{BasicEq:EM:Scalar:Maxwell1} 
yields
\Eq{
k^2D_a\A=-D_a(r^n\E)+\frac{q}{2}D_a(F^c_c-2nF).
}
Therefore (adding some constant to $\A$ in its definition, if 
necessary), we obtain
\Eq{
r^n\E = -k^2\A + \frac{q}{2}(F^c_c-2nF). 
\label{BasicEq:EM:Scalar:EbyA}
}
Thus, the gauge-invariant variables $\E$ and $\E_a$ can be expressed 
in terms of the single master variable $\A$. Finally, by inserting 
these expressions into \eqref{BasicEq:EM:Scalar:EbyEa}, we obtain 
the following wave equation for $\A$:
\Eq{
r^{n-2}D_a\left( \frac{D^a\A}{r^{n-2}} \right) -\frac{k^2}{r^2}\A
  =-r^{n-2}D_a\left(\frac{\tilde J^a}{r^{n-2}}\right)
    -\frac{q}{2r^2}(F^c_c-2nF).
\label{BasicEq:EM:Scalar:A}
}

Next, we derive an expression in terms of $\A$ for the contribution 
of the electromagnetic field to the perturbation of the 
energy-momentum tensor. First, for a scalar perturbation,  
\eqref{EMFperturbation:EMtensor} can be  written in terms of $\E$, 
$\E_a$ and the metric perturbation variables as
\Eqrsub{
&& \tau_{ab}=E_0\left( -\E+D_c(E_0 X^c)+\frac{1}{2}E_0f^c_c \right)
      g_{ab}-\frac{1}{2}E_0^2 f_{ab},\\
&& \tau^a=-E_0\left( \E^a+\frac{k}{r}E_0X^a \right),\\
&& \delta P=E_0\left(\E-D_c(E_0 X^c)-\frac{1}{2}E_0f^c_c\right),\\
&& \pi_T=0.
}
Inserting these into \eqref{Sigma_ab:def}--\eqref{Sigma_L:def} and 
using \eqref{BasicEq:EM:Scalar:EabyA} and 
\eqref{BasicEq:EM:Scalar:EbyA}, we obtain the following expressions 
for the corresponding gauge-invariant 
variables $\Sigma_{ab}$, $\Sigma_a$ and $\Sigma_L$ in terms of $\A$, 
$F$ and $F_{ab}$:
\Eqrsub{
&& \Sigma_{ab}^{EM}=\left( \frac{qk^2}{r^{2n}}\A
  +\frac{nq^2}{r^{2n}}F \right)g_{ab}
 -\frac{q^2}{2r^{2n}}F_{ab},\\
&& \Sigma_{a}^{EM}
    =-\frac{qk}{r^{2n-1}}\left(D_a\A+\tilde J_a\right),\\
&& \Sigma_{L}^{EM}= -\frac{qk^2}{r^{2n}}\A-\frac{nq^2}{r^{2n}}F. 
\label{Sigma:EMF}
}
%

\subsection{Einstein-Maxwell system: Black hole background}

For generic modes of scalar perturbations, the Einstein equations 
consist of four sets of equations of the forms 
\Eq{
E_{ab}=\kappa^2 \Sigma_{ab},\ 
E_a=\kappa^2 \Sigma_a,\ 
E_L=\kappa^2 \Sigma_L,\ 
E_T=\kappa^2 \tau_T.
}
(For the definitions of $E_{ab}, E_a, E_L$ and $E_T$, see 
Eqs.~(63)--(66) in Ref.\citen{Kodama.H&Ishibashi&Seto2000}.) If we 
introduce the variables $\tilde F^a_b$ and $\tilde F$ as
\Eq{\tilde F^a_b=r^{n-2}F^a_b,\quad
\tilde F=r^{n-2}F,
}
the equation for $E_T$ is algebraic and can be written
\Eq{
2(n-2)\tilde F + \tilde F^a_a =-S_T,
\label{STeq}
}
where
\Eq{
S_T=\frac{2r^n}{k^2}\kappa^2 \tau_T.
\label{S_T:def}}
Hence, if we introduce $X,Y$ and $Z$ defined by
\Eq{
X=\tilde F^t_t-2\tilde F,\ 
Y=\tilde F^r_r-2\tilde F,\ 
Z=\tilde F^r_t,
\label{XYZ:def}
}
as in Paper I, the original variables are expressed as
\Eqrsub{
&& \tilde F=-\frac{1}{2n}(X+Y+S_T ),\\
&& \tilde F^t_t=X+2\tilde F,\ 
   \tilde F^r_r=Y+2\tilde F,\ 
   \tilde F^r_t=Z.
\label{FbyXYZ}
}

In contrast, for the exceptional modes, \eqref{STeq} is not obtained 
from the Einstein equations. However, this equation with $S_T=0$ can 
be imposed as a gauge condition, as shown in Paper I. Under this 
gauge condition, all equations derived in this section hold without 
change. However, the variables still contain some residual gauge 
freedom, and we must eliminate it in order to extract physical 
degrees of freedom. This is done in Appendix \ref{Appendix:C}.

To obtain the expressions of the remaining Einstein equations in 
terms of $X,Y$ and $Z$, we introduce $\hat E^a_b$, $\hat E_a$ and 
$\hat E_L$ defined by 
\Eqrsubl{hat E:def}{
&& \hat E_{ab}=r^{n-2} E_{ab}
   +\frac{n(n-1)Q^2}{r^{n+2}}(F_{ab}-2n Fg_{ab}) ,\\
&& \hat E_a= r^{n-2}E_a,\\
&& \hat E_L= r^{n-2} E_L-\frac{n(n-1)Q^2}{r^{n+2}} F,
}
and separate the contributions of the electromagnetic field to the 
perturbation of the energy-momentum tensor as 
\Eq{
S_{ab}=r^{n-2}\kappa^2(\Sigma_{ab}-\Sigma_{ab}^{EM}),\ 
S_a=\frac{r^{n-1}\kappa^2}{k}(\Sigma_{a}-\Sigma_{a}^{EM}),\ 
S_L=r^{n-2}\kappa^2(\Sigma_{L}-\Sigma_{L}^{EM}).
\label{S:def}
}
Then, the remaining Einstein equations are written
\Eqrsub{
&& \hat E_{ab}   
  =\frac{k^2n(n-1)Q^2}{r^{n+2}}\frac{\A}{q}g_{ab}+S_{ab},\\
&& \hat E_a= -\frac{k n(n-1)Q^2}{r^{n+1}}
   \frac{D_a \A+\tilde J_a}{q} + \frac{k}{r}S_a,\\
&& \hat E_L
   =-\frac{k^2 n(n-1)Q^2}{r^{n+2}}\frac{\A}{q}+S_L.
}
%

As shown in Paper I, if the equations corresponding to $\hat E_a$, 
$\hat E^r_t$ and $\hat E^r_r$ are satisfied, the other equations are 
automatically satisfied, as seen from the Bianchi identities, 
provided that the energy momentum tensor satisfies the conservation 
law, which in the present case is given by the two equations%
\Eqrsubl{EMconservation:scalar:RNBH}{
&& \frac{1}{r^2}D_a(r^2 S^a) -S_L+\frac{(n-1)(k^2-nK)}{2nr^2}S_T=0,\\
&& \frac{1}{r^2}D_b(r^2S^b_a)+\frac{k^2}{r^2}S_a -n\frac{D_a r}{r}S_L
   =k^2\frac{\kappa^2 q}{r^{n+2}}\tilde J_a.
}
%

The explicit expressions of the relevant equations in terms of $X,Y$ 
and $Z$ are as follows. First, the $\hat E_a$ parts and the $\hat 
E^r_t$ part are 
\Eqrsub{
& (2r/k)\hat E_t = &-\partial_t X - \partial_r Z -\partial_t S_T,
\label{Et:RN}\\
& (2r/k)\hat E_r= &-\partial_r Y +\frac{f'}{2f} (X -Y)
       +\frac{1}{f^2}\partial_t Z 
       -\partial_r S_T + \frac{n-1}{r}S_T,
\label{Er:RN}\\
& 2\hat E^r_t=
   &f\partial_t\partial_r X 
     -\left( \frac{n-2}{r}f+ \frac{f'}{2}\right)\partial_t X
   +f\partial_t\partial_r Y 
    +\left( \frac{2f}{r}-\frac{f'}{2} \right)\partial_t Y
    \notag\\
&& +\frac{k^2}{r^2}Z
   +f\partial_t\partial_r S_T
   -\left( \frac{n-2}{r}f+\frac{f'}{2} \right)\partial_t S_T.
}
After the Fourier transformation with respect to the time coordinate 
$t$, setting $\partial_t( X,Y,Z)=-i\omega (X,Y,Z)$ and solving with 
respect to $(X',Y',Z')=\partial_r(X,Y,Z)$, we obtain 
\Eqrsubl{BasicEq:RN:XYZ}{
&& X'=\frac{n-2}{r}X
      +\left(\frac{f'}{f}-\frac{2}{r}\right)Y
    +\left(\frac{k^2}{fr^2}-\frac{\omega^2}{f^2}\right)
    \frac{Z}{i\omega}
    \notag\\
&& \qquad -2\kappa^2 E_0\left(\A'+\tilde J_r\right)
   +\left( \frac{f'}{2f}-\frac{1}{r} \right)S_T
          -\frac{2}{f}\frac{S^r_t}{i\omega}
          +2S_r,    
    \label{BasicEq:RN:X}\\
&& Y'=\frac{f'}{2f}(X-Y) +\frac{\omega^2}{f^2}\frac{Z}{i\omega}
    \notag\\
&& \qquad +2\kappa^2E_0\left(\A'+\tilde J_r\right)
    -S_T'+\frac{n-1}{r}S_T -2S_r,
    \label{BasicEq:RN:Y}\\
&& \frac{Z'}{i\omega}=X
     -2\kappa^2E_0\left(\A -\frac{\tilde J_t}{i\omega}\right)
   +S_T -\frac{2S_t}{i\omega}.
\label{BasicEq:RN:Z}
}
Next, $\hat E^r_r$ is expressed as 
\Eqr{
& 2\hat E^r_r= & 
    \frac{1}{f}\partial_t^2 X-\frac{f'}{2}\partial_r X
    +\frac{1}{f}\partial_t^2 Y
    -\left( \frac{f'}{2}+\frac{nf}{r} \right)\partial_r Y   
     +\frac{2n}{rf}\partial_t Z     
  \notag\\
&&    +\left[\frac{(n-1)(n-2)M}{r^{n+1}}
      -(2n-1)\lambda  +\frac{(n-1)Q^2}{r^{2n}} \right]X 
   \notag\\
&&  +\left[\frac{k^2-nK}{r^2}-\frac{(n+1)(n-2)M}{r^{n+1}}
     +(n+1)\lambda +\frac{(2n^2-2n-1)Q^2}{r^{2n}} \right]Y
   \notag\\
&&
   +\frac{1}{f}\partial_t^2 S_T 
   -\left( \frac{f'}{2}+\frac{nf}{r} \right)\partial_r S_T
   \notag\\
&& +\left[\frac{k^2+n(n-2)K}{r^2}-\frac{(n+1)(n-2)M}{r^{n+1}}
     \right.\notag\\
&&\quad\qquad \left.
   -(n^2-1)\lambda + \frac{(n^2-n+1)Q^2}{r^{2n}} \right] S_T.
}
Applying the Fourier transformation with respect to time to the 
corresponding equation and eliminating $X',Y'$ and $Z'$, with the 
help of \eqref{BasicEq:RN:XYZ}, we obtain the following linear 
constraint on $X,Y$ and $Z$: 
\Eqr{
&& 
  \left[\omega^2r^2
     +K\left(\lambda r^2+\frac{n(n-1)M}{r^{n-1}} 
       -\frac{(n-1)(2n-1)Q^2}{r^{2n-2}}\right) \right.\notag\\
&&\quad   -\frac{M}{r^{n-1}}\left( 
       n(n+1)\lambda r^2+\frac{(n^2-1)M}{r^{n-1}}
      -\frac{3n(n-1)Q^2}{r^{2n-2}} \right) \notag\\
&&\quad\left. +\frac{Q^2}{r^{2n-2}}\left( 
       n(2n-1)\lambda r^2 -\frac{n(n-1)Q^2}{r^{2n-2}} \right) 
       \right]X
   \notag\\
&& \quad +\left[ \omega^2r^2 -k^2 f 
   + nK^2-(n-1)K\lambda r^2
   -\frac{4KnM}{r^{n-1}}
   -\frac{K(n^2-4n+1)Q^2}{r^{2n-2}} \right.\notag\\
&&\quad \left.  +\frac{M}{r^{n-1}}\left( 
   \frac{(n+1)^2M}{r^{n-1}}-\frac{4nQ^2}{r^{2n-2}} \right)
   +\frac{Q^2}{r^{2n-2}}\left( n(n-1)\lambda r^2
    +\frac{nQ^2}{r^{2n-2}} \right)
   \right](Y+S_T) 
   \notag\\
&& \quad -\frac{1}{r}\left[ 
    n \omega^2 r^2
   +\left(\lambda r^2 -\frac{(n-1)M}{r^{n-1}}
      +\frac{(n-1)Q^2}{r^{2n-2}}\right)k^2\right]
   \frac{Z}{i\omega} \notag\\
&& \quad +2\kappa^2E_0 f\left[nrf(\A'+\tilde J_r)+k^2\A  \right]
     \notag\\
&& \quad -2nrf^2 S_r -r^2f'\frac{S^r_t}{i\omega}
   +2r^2fS^r_r
  =0.
\label{BasicEq:RN:constraint}
}
%

Using almost the same method as in Paper I, we can reduce this 
constrained system for $X,Y$ and $Z$ to a single second-order ODE 
with source. The master variable in the present case is given by%
\Eq{
\Phi=\frac{n}{r^{n/2-2}H}
  \left( \frac{\tilde F^r_t}{i\omega r}+2\tilde F \right) 
  =-\frac{X+Y+S_T-\frac{nZ}{i\omega r}}{r^{n/2-2}H},
\label{RN:scalar:MasterVar:metric}
}
where 
\Eqr{
&& H=m +\frac{n(n+1)}{2}x - n^2z,\\
&& m=k^2-nK,\\
&& x=\frac{2M}{r^{n-1}},\ 
   y=\lambda r^2,\ 
   z=\frac{Q^2}{r^{2n-2}}.
}
This master variable coincides with that for the neutral and 
source-free case in Paper I for $Q=0$ and $S_T=0$. The master 
equation for $\Phi$ is %
\Eq{
f(f\Phi')' +(\omega^2-V_S)\Phi=S_\Phi.
\label{RN:scalar:MasterEq:metric}
}
Here, the effective potential $V_S$ is expressed as
\Eqr{
&& V_S=\frac{f(r)U_S(r)}{16r^2H^2};\\
&& U_S=\left[-n^3 (n+2) (n+1)^2 x^2+4n^2(n+1)\{n(n^2+6 n-4) z
      +3(n-2) m\} x \right.\notag\\
&&\quad\left. -12 n^5 (3 n-2) z^2-8 n^2 (11 n^2-26 n+12) m z
      -4 (n-2) (n-4) m^2\right] y \notag\\
&&\quad  +n^4 (n+1)^2 x^3+n(n+1)\left\{-3 n^2 (5 n^2-5 n+2) z  
    +4 (2 n^2-3 n+4) m  \right.\notag\\
&&\quad \left. +n (n-2) (n-4) (n+1) K\right\} x^2
    +4n\left[n^2 (4 n^3+5 n^2-10 n+4) z^2 \right.\notag\\
&&\quad -\left\{n (34-43 n+21 n^2) m
        + n^2 (n+1) (n^2-10 n+12) K\right\} z
  \notag\\
&&\quad \left.-3 (n-4) m^2-3 n (n+1) (n-2) K m\right] x
        -4 n^5 (3 n-2) z^3
\notag\\
&&\quad +12n^2\left\{2(-6 n+3 n^2+4) m+n^2 (3 n-4) (n-2) K\right\} z^2
\notag\\ 
&&\quad +\left\{4 (13 n-4) (n-2) m^2
                 +8 n^2 (11 n^2-18 n+4) K m\right\} z
\notag\\
&&\quad +16 m^3+4 n (n+2) K m^2,
}
and the source term $S_\Phi$ has the following structure:
\Eqr{
&& S_\Phi=\frac{f}{r^{n/2}H}\left[ \kappa^2 E_0\left( 
      \frac{P_{S1}}{H}\left(\A-\frac{\tilde J_t}{i\omega}\right)
      +2nrf \tilde J_r 
   +2k^2\frac{\tilde J_t}{i\omega}
   +2nf\frac{r\partial_r \tilde J_t}{i\omega} \right) \right.
   \notag\\
&& \qquad\qquad\qquad
   -HS_T -\frac{P_{S2}}{H}\frac{S_t}{i\omega }
   -2nf\frac{r\partial_r S_t}{i\omega}
   -2nrfS_r \notag\\
&& \qquad\qquad\qquad \left.
   +\frac{P_{S3}}{H}\frac{rS^r_t}{i\omega}
   +2r^2\frac{\partial_r S^r_t}{i\omega}
   +2r^2S^r_r
   \right],
\label{RN:scalar:MasterEq:Source:Phi}
}
where $P_{S1}, P_{S2}$ and $P_{S3}$ are functions of $r$ expressed 
as polynomials of $x,y$ and $z$. Their explicit expressions are given 
in Appendix \ref{Appendix:B}.  

The basic variables $X,Y$ and $Z$ are expressed in terms of the 
master variable $\Phi$ as 
\Eqrsubl{XYZbyPhi:RNBH}{
&& X=r^{n/2-2}\left[ \left( \frac{\omega^2r^2}{f}
     -\frac{P_{X0}}{16H^2}\right)\Phi
     +\frac{P_{X1}}{4H}r\partial_r\Phi \right] +X_s,\\
&& Y=r^{n/2-2}\left[ \left( -\frac{\omega^2r^2}{f}
     -\frac{P_{Y0}}{16H^2}\right)\Phi
     +\frac{P_{Y1}}{4H}r\partial_r\Phi \right] +Y_s,\\
&& Z=i\omega r^{n/2-1}\left[ \frac{P_Z}{4H}\Phi
        -fr\partial_r\Phi \right] +Z_s,
}
where $X_s,Y_s$ and $Z_s$ are contributions from the source terms given by
\Eqrsub{
& X_s=& \kappa^2 E_0\left( \frac{P_{XA}}{2H^2}\A
         -\frac{2nrf}{H}(\partial_r\A+\tilde J_r)
     - \frac{n P_{X2}}{2H^2}\frac{\tilde J_t}{i\omega} \right)
   \notag\\
&& +\frac{n P_{X2}}{2H^2}\frac{S_t}{i\omega }
    +\frac{2nrf}{H} S_r 
   +\frac{nP_{X3}}{H^2}
   \frac{rS^r_t}{i\omega}-\frac{2r^2}{H}S^r_r,\\
& Y_s=& \kappa^2 E_0\left(\frac{P_{YA}}{2H^2}\A
         +\frac{2nrf}{H}(\partial_r\A+\tilde J_r)
   -\frac{nP_{Y2}}{2H^2}\frac{\tilde J_t}{i\omega}\right)\notag\\
&& + \frac{nP_{Y2}}{2H^2}\frac{S_t}{i\omega }
    -\frac{2nrf}{H}S_r 
    +\frac{nP_{Y3}}{H^2}
   \frac{rS^r_t}{i\omega}+\frac{2r^2}{H}S^r_r -S_T,\\
&Z_s=& \kappa^2 E_0 \frac{2nrf}{H}(-i\omega \A
       +\tilde J_t )
 -\frac{2nrf}{H}S_t+\frac{2r^2}{H}S^r_t.
}
Here, $P_{X0},P_{X1},P_{Y0},P_{Y1},P_{Z},P_{XA},P_{X2},P_{YA}$ and 
$P_{Y2}$ are functions of $r$ and are given in Appendix 
\ref{Appendix:B}. 

In particular, $\tilde F$ is expressed as
\Eq{
\tilde F=r^{n/2-2}\left[\frac{4H^2-nP_Z}{8nH}\Phi
  +\frac{rf}{2}\partial_r\Phi \right] -\frac{Z_s}{2i\omega r}.
}
Hence, using the relation
\Eq{
\tilde F^a_a-2n\tilde F=-4(n-1)\tilde F -S_T,
}
we can rewrite the Maxwell equation \eqref{BasicEq:EM:Scalar:A} as
\Eqr{
&& r^{n-2}D_a\left( \frac{D^a\A}{r^{n-2}} \right) 
   -\frac{1}{r^2}\left(k^2
        +\frac{2n^2(n-1)^2zf}{H}\right)\A 
\notag\\
&& \qquad\qquad    
   =\frac{(n-1)q}{r^{n/2+2}}\left( \frac{4H^2-nP_Z}{4nH}\Phi
      +fr\partial_r\Phi \right) 
    +\frac{S_A}{f},
\label{RN:scalar:MasterEq:EMF}}
where
\Eq{
S_A= -\left( \frac{2n^2(n-1)^2zf^2}{r^2H}+\omega^2 \right)
        \frac{\tilde J_t}{i\omega}
        -r^{n-2}f\partial_r\pfrac{f\tilde J_r}{r^{n-2}}
        +\frac{2(n-1)E_0}{i\omega H}f\left( nf S_t
        -r S^r_t \right).
\label{RN:scalar:MasterEq:Source:A}
}

Thus, the two coupled second-order ODEs 
\eqref{RN:scalar:MasterEq:metric} and \eqref{RN:scalar:MasterEq:EMF} 
represent the master equations with a source for a scalar 
perturbation of the Einstein-Maxwell system in the background 
\eqref{metric:GSBH}. 

Finally, we note that in terms of the variable
\Eq{
\tilde \Omega:=r^{n/2}H \Phi,
}
\eqref{XYZbyPhi:RNBH} can be rewritten as 
\Eqrsubl{FbyTPhi:covariant}{
&&\tilde F  =\frac{1}{4nr^2H}\left[ (2H-nrf')\tilde\Omega
              +2nr Dr\cdot D\tilde\Omega \right]             
               -\frac{1}{i\omega H}(-nf\tilde S_t+rS^r_t),\\ 
&& \tilde F_{ab}+(n-2)\tilde F g_{ab}
   =\frac{1}{H}\left( D_aD_b\tilde\Omega
        -\frac{1}{2}\Box\tilde\Omega g_{ab} \right)
   +\tilde S^a_b,
}
where 
\Eq{
\tilde S_a=S_a-\kappa^2 E_0(D_a\A+\tilde J_a),
}
and 
\Eqr{
& \tilde S_{ab}=& -\frac{2nr}{H}
    \left( D_a r \tilde S_b+D_b r\tilde S_a
          -\frac{2D_a r D_b r}{f}D_c r\tilde S^c \right) \notag\\
&& -\frac{1}{i\omega}{nf}{rH}\left( g_{ab} 
            -\frac{2}{f}D_a r D_b r\right)
            \epsilon^{cd}D_c(r^2\tilde S_d) \notag\\
&& +\frac{2r^2}{H}S_{ab}
   -\frac{r^2}{H}\left( S^c_c+\frac{1}{2}S_T \right)g_{ab}.
}
These equations are formal, since they contain the factor 
$i/\omega$, which is equivalent to integration with respect to the 
time coordinate $t$. However, when $S_a=\kappa^2 E_0\tilde J_a$ and 
$S^r_t=0$, this factor can be eliminated through the replacements 
$\tilde S_t \tend i\omega \kappa^2E_0\A$ and 
$\epsilon^{ab}D_a(r^2\tilde S_b) \tend -i\omega (n-2)\kappa^2r^2E_0 
\A$. In this case, 
\eqref{FbyTPhi:covariant} provides expressions that are manifestly 
covariant as equations in the 2-dimensional spacetime $\N^2$. The 
source term $S_\Phi$ for $\Phi$ can also be rewritten in such a 
covariant form with the help of the energy-momentum conservation 
laws 
\eqref{EMconservation:scalar:RNBH}: 
\Eqr{
& S_\Phi =& \frac{f}{r^{n/2} H}\left[ 4k^2 \kappa^2 E_0 \A 
  + \frac{i}{\omega}2nH^2r^{n-1}\epsilon^{ab}D_a
   \left( \frac{r^{n-2}f}{H^2} \tilde S_b \right) 
   \right.\notag\\
&& \left. +\frac{i}{\omega}\frac{2(nH+2rH')}{H^2}rS^r_t
   +2r^2S^a_a -H S_T\right].
}
The comment made above regarding covariance applies to this 
expression as well as to \eqref{RN:scalar:MasterEq:EMF}. 

\subsection{Reduction to decoupled equations}

As in the vector perturbation case, we can transform the master 
equations \eqref{RN:scalar:MasterEq:metric} and 
\eqref{RN:scalar:MasterEq:EMF} into two decoupled second-order ODEs 
by introducing new master variables written as linear combinations 
of $\Phi$ and $\A$. The only difference in this case is that the 
coefficients are not constant. These new variables are given by 
\Eq{
\Phi_\pm=a_\pm \Phi + b_\pm \A,
}
with 
\Eqrsub{
&& (a_+,b_+)=\left(\frac{m}{n}Q+\frac{(n+1)(M+\mu)}{2r^{n-1}}Q,
    \frac{(n+1)(M+\mu)Q}{qr^{n/2-1}}\right),\\
&& (a_-,b_-)=\left((n+1)(M+\mu)-\frac{2nQ^2}{r^{n-1}},
    -\frac{4nQ^2}{qr^{n/2-1}} \right),
}
where $\mu$ is a positive constant satisfying
\Eq{
\mu^2=M^2+\frac{4mQ^2}{(n+1)^2}.
}

The reduced master equations have the canonical forms
\Eq{
f(f\Phi_\pm')'+\left(\omega^2-V_{S\pm}\right)\Phi_\pm
  =S_{S\pm}.
}
Here, if we define the parameter $\delta$ by
\Eq{
\mu=(1+2m\delta)M,
}
the effective potentials $V_{S\pm}$ are expressed as
\Eq{
V_{S\pm} =\frac{fU_\pm}{64r^2H_\pm^2},
}
with
\Eqrsub{
&& H_+=1-\frac{n(n+1)}{2}\delta x,\\
&& H_-=m+\frac{n(n+1)}{2}(1+m\delta)x,
}
and 
\Eqrsub{
& U_+ =
& \left[-4 n^3 (n+2) (n+1)^2 \delta^2 x^2-48 n^2 (n+1) (n-2) \delta x
\right.\notag\\
&&\left.   -16 (n-2) (n-4)\right] y
  -\delta^3 n^3 (3 n-2) (n+1)^4 (1+m \delta) x^4
\notag\\
&&   +4 \delta^2 n^2 (n+1)^2 
   \left\{(n+1)(3n-2) m \delta+4 n^2+n-2\right\} x^3
\notag\\   
&&   +4 \delta (n+1)\left\{
   (n-2) (n-4) (n+1) (m+n^2 K) \delta-7 n^3+7 n^2-14 n+8
   \right\}x^2
\notag\\   
&&  + \left\{16 (n+1) \left(-4 m+3 n^2(n-2) K\right) \delta
     -16 (3 n-2) (n-2) \right\}x
\notag\\   
&&    +64 m+16 n(n+2) K,\\
& U_- =
  & \left[-4 n^3 (n+2) (n+1)^2 (1+m \delta)^2 x^2
      +48 n^2 (n+1) (n-2) m (1+m \delta) x  \right.
\notag\\
&& \left.  -16 (n-2) (n-4) m^2\right] y
     -n^3 (3 n-2) (n+1)^4 \delta (1+m \delta)^3 x^4
\notag\\
&& -4 n^2 (n+1)^2 (1+m \delta)^2 
     \left\{(n+1)(3 n-2) m \delta-n^2\right\} x^3
\notag\\  
&&  +4 (n+1) (1+m \delta)\left\{ m (n-2) (n-4) (n+1) (m+n^2 K) \delta
  \right. \notag\\
&& \left. \quad  +4 n (2 n^2-3 n+4) m+n^2 (n-2) (n-4) (n+1)K \right\}x^2
\notag\\
&&  -16m \left\{ (n+1) m \left(-4 m+3 n^2(n-2) K\right) \delta
\right.\notag\\
&&\left.  +3 n (n-4) m+3 n^2 (n+1) (n-2)K \right\}x
\notag\\
&&      +64 m^3+16 n(n+2)m^2 K.
}
The source terms $S_{S\pm}$ are  linear combinations of $\bar 
S_\Phi=S_\Phi|_{\A=0}$ given in 
\eqref{RN:scalar:MasterEq:Source:Phi} and $S_\A$ given in  
\eqref{RN:scalar:MasterEq:Source:A}:
\Eq{
S_{S\pm}=a_\pm  \bar S_\Phi + b_\pm  S_\A.
}

Here, note that the following relations hold:
\Eqr{
&& Q^2=(n+1)^2M^2 \delta( 1+m\delta),\\
&& H=H_+ H_-.
}
{}From these relations, we find that $Q=0$ corresponds to 
$\delta=0$, and in this limit, $\Phi_-$ coincides with $\Phi$, and 
its equation coincides with the mater equation for the master 
variable $\Phi$ derived in Paper I. Hence, $\Phi_-$ and $\Phi_+$ 
represent the gravitational mode and the electromagnetic mode, 
respectively. For $n=2, K=1$ and $\lambda=0$, these variables 
$\Phi_+$ and $\Phi_-$ are proportional to the variables for the 
polar modes, $Z^{(+)}_1$ and $Z^{(+)}_2$, appearing in Ref. 
\citen{Chandrasekhar.S1983B}, and $V_{S+}$ and $V_{S-}$ coincide 
with the corresponding potentials, $V^{(+)}_1$ and $V^{(+)}_2$, 
respectively.

\subsection{Einstein-Maxwell system: Nariai background}

We can derive the master equations for the Einstein-Maxwell system 
in the Nariai-type background \eqref{metric:Nariai} in almost the 
same way as in the black hole case. Actually, the calculations are 
much simpler in this case. For this reason, we give only key 
equations. As in the previous case, the elimination of the residual 
gauge freedom for the exceptional modes is discussed in Appendix 
\ref{Appendix:C}. 

First, by defining the variables $X,Y$ and $Z$ by \eqref{XYZ:def}, 
after the Fourier transformation, the equations corresponding to 
$\hat E_\rho,\hat E_t$ and $\hat E^\rho_t$ can be written as
\Eqrsub{
&& X'= \frac{f'}{f}Y
    +\left(\frac{k^2}{fa^2}-\frac{\omega^2}{f^2}\right)
    \frac{Z}{i\omega}
    \notag\\
&& \qquad -2\kappa^2 E_0\left(\A'+\tilde J_\rho\right)
   +\frac{f'}{2f}S_T
          -\frac{2}{f}\frac{S^\rho_t}{i\omega}
          +2S_\rho,    
    \label{BasicEq:Nariai:X}\\
&& Y'=\frac{f'}{2f}(X-Y) +\frac{\omega^2}{f^2}\frac{Z}{i\omega}
    \notag\\
&& \qquad +2\kappa^2E_0\left(\A'+\tilde J_\rho\right)
    -S_T' -2S_\rho,
    \label{BasicEq:Nariai:Y}\\
&& \frac{Z'}{i\omega}=X
     -2\kappa^2E_0\left(\A -\frac{\tilde J_t}{i\omega}\right)
   +S_T -2\frac{S_t}{i\omega}.
\label{BasicEq:Nariai:Z}
}
The constraint equation obtained from the $E^\rho_\rho$ equation is 
\Eqr{
&& (\omega^2+\sigma)X
   +\left(\omega^2+\sigma-\frac{k^2}{a^2}f\right)(Y+S_T)
   -\frac{k^2\sigma \rho}{a^2}\frac{Z}{i\omega}
   \notag\\
&& +\frac{2\kappa^2E_0 k^2f}{a^2}\A
   +2\sigma \rho \frac{S^\rho_t}{i\omega}
   +2fS^\rho_\rho=0.
}
%
These equations can be reduced to a single second-order ODE for $F$, 
\Eq{
f\frac{d}{d\rho}\left(f\frac{dF}{d\rho}\right) +(\omega^2-V_F)F=S_F,
\label{Nariai:scalar:MasterEq:metric}
}
where 
\Eqr{
&& V_F=f\left( \frac{k^2}{a^2}-2\sigma \right),\\
&& S_F= \frac{k^2 f}{2na^n}\left[2\kappa^2E_0\left( 
        2\A - \frac{\tilde J_t}{i\omega} \right)
    -S_T+2\frac{S_t}{i\omega}
    +\frac{2a^2}{k^2}\frac{\partial_\rho S^\rho_t}{i\omega}
    +\frac{2a^2}{k^2}S^\rho_\rho   \right].
}
Hence, in the present case, we can use $F$, which is related to $X$ and $Y$ by
\Eq{
F:=-\frac{X+Y+S_T}{2na^{n-2}},
\label{Nariai:scalar:MasterVar:metric}
}
as the master variable. 
$X,Y$ and $Z$ are expressed in terms of $F$ as
\Eqrsub{
&X= & \frac{2na^n}{k^2}\left[\left( \frac{\omega^2}{f}
      -\frac{k^2}{a^2}+\sigma \right)F
      -\sigma \rho \partial_\rho F\right]
   -2\kappa^2E_0\A -\frac{2a^2}{k^2}S^\rho_\rho,\\
&Y= & \frac{2na^n}{k^2}\left[-\left(\frac{\omega^2}{f}+\sigma \right)F
     +\sigma \rho\partial_\rho \right]  +2\kappa^2E_0 \A
   +\frac{2a^2}{k^2}S^\rho_\rho -S_T,\\
&Z= & -i\omega\frac{2na^n}{k^2}
   \left( \sigma \rho F+f\partial_\rho F \right)
       +\frac{2a^2}{k^2}S^\rho_t.
}
%

Finally, the Maxwell equation \eqref{BasicEq:EM:Scalar:A} also holds 
in the present case, in which it becomes 
\Eq{
f\frac{d}{d\rho}\left( f\frac{d\A}{d\rho} \right)
  +\left( \omega^2-\frac{k^2}{a^2}f \right)\A
  =2(n-1)a^{n-2}f E_0 F + S_\A,
}
where
\Eq{
S_\A=  fE_0S_T-i\omega\tilde J_t-f^2\partial_\rho\tilde J_\rho.
}
By taking linear combinations of this equation and 
\eqref{Nariai:scalar:MasterEq:metric} and introducing the variables 
\Eq{
\Phi_\pm=a_\pm F + b_\pm \A,
}
we obtain the decoupled equations
\Eq{
f\frac{d}{d\rho}\left(f \frac{d\Phi_\pm}{d\rho} \right)
 +\left[ \omega^2-\left( \frac{k^2}{a^2}\pm \mu -\sigma \right)f   
       \right]\Phi_\pm =S_\pm,
}
with 
\Eq{
S_\pm=a_\pm\bar S_F + b_\pm S_\A,
}
where $\bar S_\Phi=S_\Phi|_{\A=0}$, $\mu$ is a positive constant satisfying
\Eq{
\mu^2=\sigma^2+\frac{4(n-1)^2Q^2}{a^{2n+2}}k^2,
}
and
\Eqrsub{
&& (a_+,b_+) = \left( \frac{2(n-1)Q}{a^{2}},
                     (\mu+\sigma)\frac{Q}{q} \right),\\
&& (a_-,b_-) = \left( \mu+\sigma,
              -\frac{2(n-1)k^2Q^2}{qa^{2n}} \right).
}
%

\section{Stability of Black Holes}

\begin{figure}[t]
\begin{minipage}{\halftext}
\centerline{\includegraphics[width=6cm]{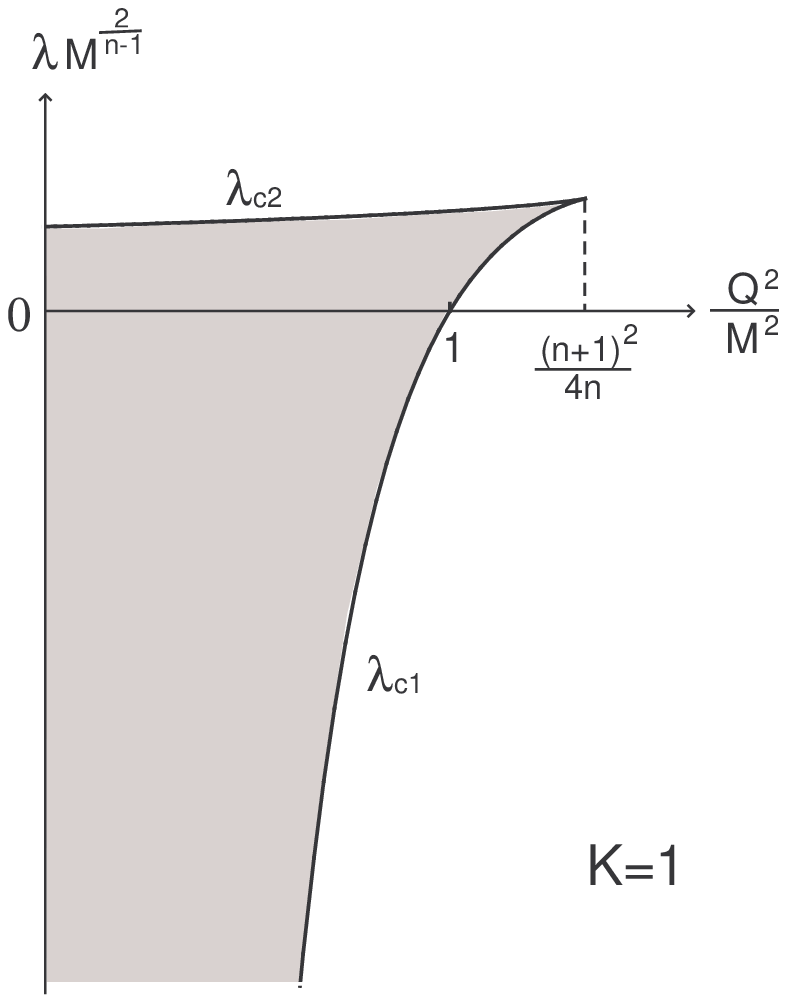}}
\end{minipage}
\begin{minipage}{\halftext}
\centerline{\includegraphics[width=6cm]{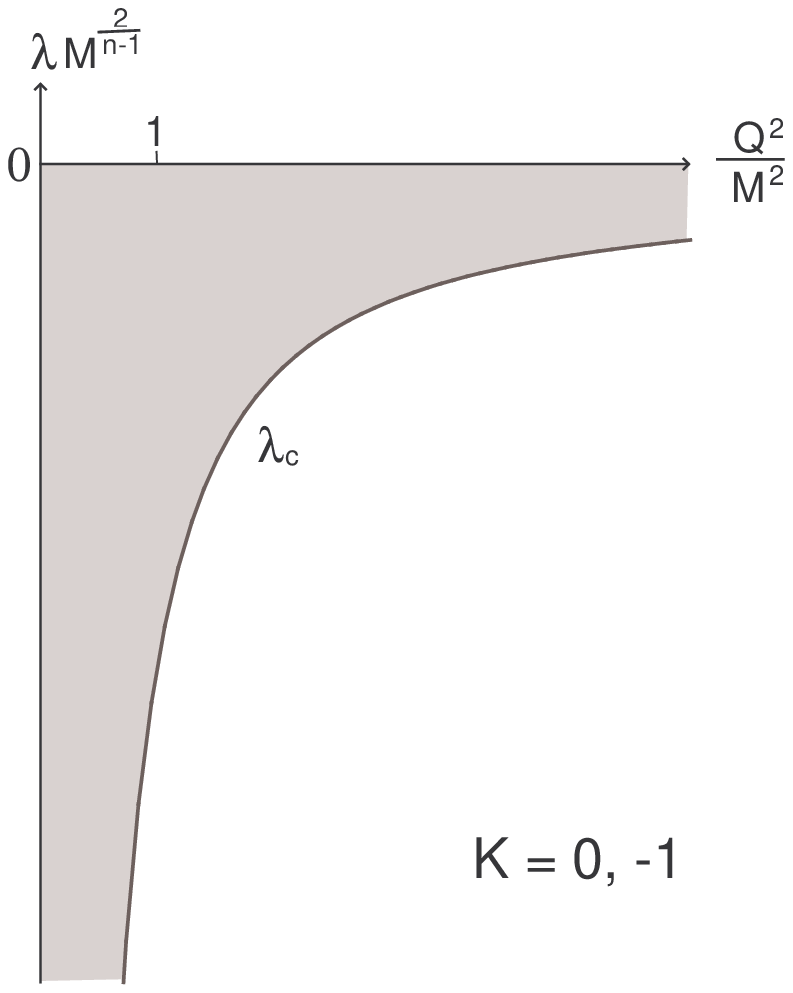}}
\end{minipage}
\caption{The parameter ranges in which the spacetime contains a 
regular black hole (the shaded regions). }\label{fig:BHcondition}
\end{figure}

In this section, we consider what we are able to conclude at this 
time about the stability of generalised static black holes with 
charge from the results of our formulation. In the present paper, we 
consider only the stability in the static region outside the black 
hole horizon with respect to a perturbation whose support is compact 
on the initial surface. This region is 
represented as $r>r_H$ for $\lambda\le0$ and $r_H<r<r_c$ for 
$\lambda>0$. As mentioned in \S\ref{subsec:BG}, such a region exists 
only for restricted ranges of the parameters $M,Q$ and $\lambda$. 
These parameter ranges are shown in Fig. \ref{fig:BHcondition} (see 
Appendix \ref{Appendix:D} for details). 

To study the stability in this region, as in Paper II, we utilize 
the fact that for any perturbation type, the perturbation equations 
in the static region of the spacetime are reduced to an eigenvalue 
problem of the type
\Eq{
\omega^2 \Phi = A\Phi,
}
where $A$ is a self-adjoint operator, 
\Eq{
A = -\frac{d^2}{dr_*^2} + V(r); \quad dr_*=\frac{dr}{f},
}
with $V(r)$ being equal to $V_T(r)$, $V_{V\pm}(r)$ or $V_{S\pm}(r)$. 
We regard the black hole to be stable if the spectrum of $A$, i.e., 
$\omega^2$, is non-negative.

To be precise, we must specify a boundary condition for $\Phi$ at 
$r=\infty$ in the case $\lambda<0$, for which the range of $r_*$ has 
an upper bound. In the present paper, we adopt the simplest 
condition, $\Phi\tend 0$ as $r \tend \infty$, in this case, which 
corresponds to the Friedrichs extension of $A$. For $\lambda\ge0$, 
the range of $r_*$ is $(-\infty,+\infty)$ and the operator $A$ is 
essentially self-adjoint; i.e., it has a unique self-adjoint 
extension\cite{Ishibashi.A&Kodama2003A}.

In general, if $\Phi(r)$ is a function with compact support 
contained in $r>r_H$ (or $r_H<r<r_c$ for $\lambda>0$), we can 
rewrite the expectation value of $A$, $(\Phi,A\Phi)$, as
\Eq{
(\Phi,A\Phi)=\int dr_*\left( \left|\frac{d\Phi}{dr_*}\right|^2 
  +V|\Phi|^2 \right).
\label{EVofA}
}
For the Friedrichs extension of $A$, the lower bound of this 
quantity coincides with the lower bound of the spectrum of 
$A$ with domain $C^\infty_0(r_*)$. 
Hence, if we can show that the right-hand side of \eqref{EVofA} 
is non-negative, then we can conclude that the system is stable. In 
particular, if $V$ is non-negative, this condition is trivially 
satisfied. However, such a lucky situation is not realized in most 
cases. One powerful method that can be used to show the positivity 
of $A$ beyond such a simple situation is to deform the right-hand 
side of \eqref{EVofA} by partial integration in terms of a function 
$S$ as 
\Eq{
(\Phi,A\Phi)=\int dr_* \left( |\tilde D\Phi|^2+\tilde V|\Phi^2| 
\right),
}
where 
\Eqr{
&& \tilde D=\frac{d}{dr_*}+S,\\
&& \tilde V= V + f\frac{dS}{dr} -S^2.
}
We call this procedure {\em the $S$-deformation of $V$} in this 
paper. As in Paper II, this is the main tool for the analysis in the 
present paper.

\subsection{Tensor perturbation}

If we apply the $S$-deformation with 
\Eq{
S=-\frac{nf}{2r}
}
to \eqref{VT:GSBH}, we obtain 
\Eq{
\tilde V_T =\frac{f}{r^2}\left[\lambda_L-2(n-1)K \right],
}
irrespective of the $r$-dependence of $f(r)$. 
Hence, the effective potential $\tilde V_T$ is positive, and the 
system is perturbatively stable with respect to a tensor 
perturbation if 
\Eq{
\lambda_L\ge 2(n-1)K.
\label{StabilityCondition:Tensor}
}
In particular, this guarantees the stability of maximally symmetric 
black holes for $K=1$ and $K=0$, since $\lambda_L$ is related to the 
eigenvalue $k_T^2$ of the positive operator $-D\cdot D$ as 
$\lambda_L=k_T^2+2nK$ when $\K^n$ is maximally symmetric. In 
contrast, even in the maximally symmetric case, $\tilde V_T$ becomes 
negative in the range $0<k_T^2<2$ for $K=-1$, and we cannot conclude 
anything about the stability for $K=-1$ from this argument alone. 

Note that the condition \eqref{StabilityCondition:Tensor} is just a 
sufficient condition for stability, and it is not a necessary 
condition in general. In fact, for a tensor perturbation, we can 
obtain stronger stability conditions directly from the positivity of 
$V_T$ if we restrict the range of parameters. For example, for $K=1$ 
and $\lambda=0$, it is easy to see that $V_T$ is positive if
\Eq{
\lambda_L+2-2n+\frac{n(n-1)\sqrt{M^2-Q^2}}{M+\sqrt{M^2-Q^2}}\ge 0
}
for $M^2\ge Q^2>8n(n-1)M^2/(3n-2)^2$, and 
\Eq{
\lambda_L+\frac{n^2-10n+8}{4}\ge 0
\label{StabilityCond:Tensor:smallQ}}
for $Q^2\le 8n(n-1)M^2/(3n-2)^2$. Thus, if we do not restrict the 
range of $Q^2$, we obtain the same sufficient condition for 
stability as \eqref{StabilityCondition:Tensor}, but for the 
restricted range $Q^2\le 8n(n-1)M^2/(3n-2)^2$, we obtain the 
stronger sufficient condition \eqref{StabilityCond:Tensor:smallQ}, 
which coincides with the condition obtained in Paper II for the case 
$Q=0$.

\begin{figure}[t]
\begin{minipage}{\halftext}
\includegraphics[width=6cm]{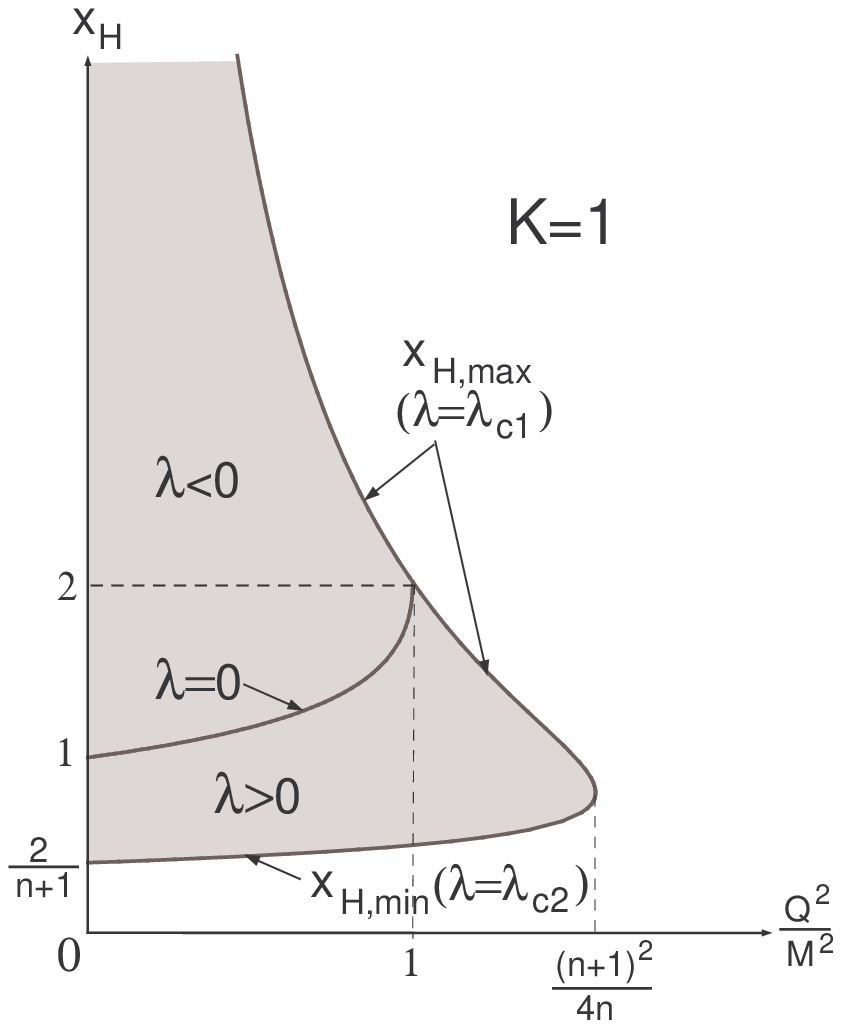}
\end{minipage}
\hspace{1cm}
\begin{minipage}{\halftext}
\includegraphics[width=5cm]{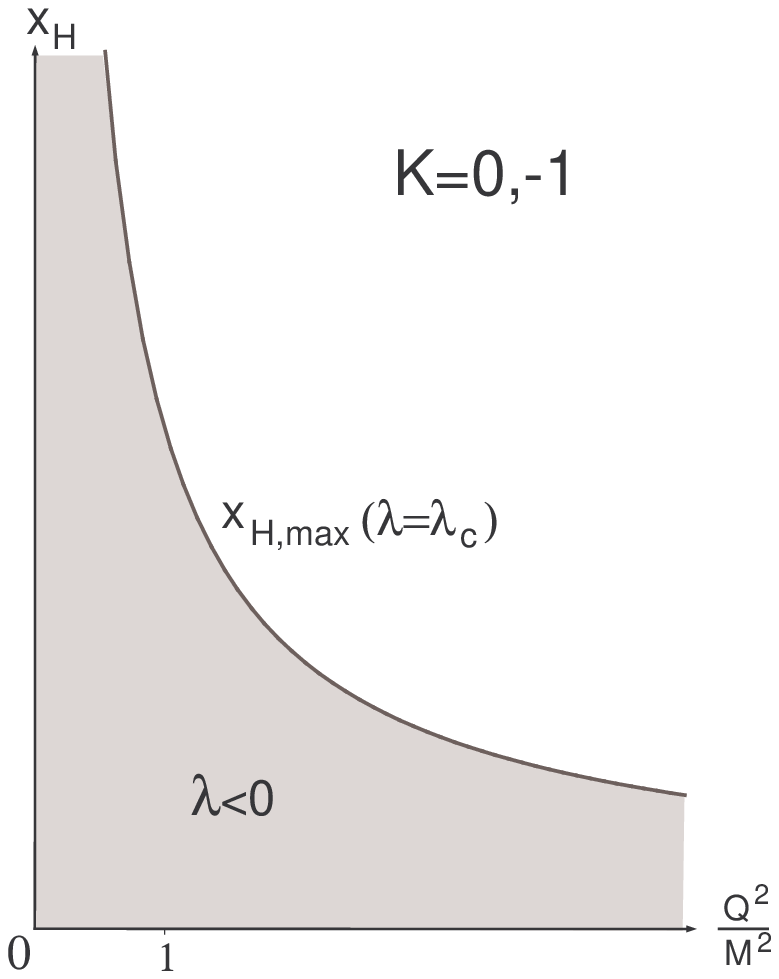}
\end{minipage}
\caption{Ranges of the value $x_H$.}
\label{fig:xh}
\end{figure}

Similarly, for $K=-1$ and $\lambda<0$, rewriting $V_T$ as
\Eq{
V_T=\frac{f}{r^2}\left(\lambda_L-(3n-2)K+ \frac{n(n+2)}{4}f
   +\frac{n(n+1)M}{r^{n-1}}-\frac{n^2Q^2}{r^{2n-2}} \right),
}
we obtain a sufficient condition for stability stronger than 
\eqref{StabilityCondition:Tensor}, 
\Eq{
\lambda_L+3n-2=k_T^2+n-2\ge0,
}
from $V_T>0$ if we restrict the range of $\lambda$ to
\Eq{
\lambda\ge -\pfrac{(n+1)M}{nQ^2}^{\frac{2}{n-1}}
  \left( 1+\frac{(n^2-1)M^2}{n^2Q^2} \right).
}
This condition is sufficient to guarantee the stability of a 
maximally symmetric black hole with $K=-1$ for $n\ge2$. However, if 
we extend the range of $\lambda$ to the whole allowed range, i.e., 
that satisfying $\lambda\ge \lambda_{c-}$, 
\eqref{StabilityCondition:Tensor} is the strongest condition that 
can be obtained only from $V_T>0$.

\subsection{Vector perturbation}

For $V=V_{V\pm}$ in \eqref{Vpm:Vector} and 
\Eq{
S=\frac{nf}{2r},
}
we obtain
\Eqr{
&& \tilde V_{V\pm} =\frac{f}{r^2}\left[ m_V
   +\frac{(n^2-1)M\pm\Delta}{r^{n-1}} \right],\\
&& m_V=k_V^2-(n-1)K.
}
Hence, from $k_V^2\ge(n-1)K$, we find that $\tilde V_{V+}$ is always 
positive, and  static charged black holes are stable with respect to 
the electromagnetic mode of the vector perturbation. In contrast, 
from
\Eqr{
&& \tilde V_{V-}=\frac{f}{r^2}
       \frac{m_V h}{(n^2-1)M+\Delta},\\
&& h:=(n^2-1)M-\frac{2n(n-1)Q^2}{r^{n-1}}+\Delta,
}
it is seen that $\tilde V_{V-}$ may become negative: Because $h$ is 
a monotonically increasing function of $r$, $\tilde V_{V-}$ is 
positive if and only if $h(r_H)\ge0$. 

\subsubsection{The case $\lambda\ge0$} In this case, the background 
spacetime contains a regular black hole only for $K=1$, and the 
static region outside the black hole is given by 
$r_H<r<r_c$ $(\le+\infty)$. In this region, from \eqref{xHmax}, we 
have $2M/r^{n-1}<x_H\le x_{H,{\rm max}}\le (n+1)M^2/(nQ^2)$. Hence, 
$h>0$, and the black hole is stable in this case. 

\subsubsection{The case $\lambda<0$} In this case, as shown in 
Appendix \ref{Appendix:D}, the spacetime contains a regular black 
hole if $\lambda\ge\lambda_c$ and $2M/r^{n-1}<x_H\le x_{H,{\rm 
max}}$. Hence, from \eqref{xHmax}, we obtain the relation 
\Eq{
h\ge \sqrt{(n^2-1)^2M^2+2n(n-1)m_V Q^2}
    -\sqrt{(n^2-1)^2M^2-4Kn(n-1)^2Q^2}.
}
For $K=0,1$, $h>0$ follows from this. Hence, the black hole is 
stable. In contrast, for $K=-1$, the right-hand side of this 
inequality becomes negative for $k_V^2<n-1$. Hence, $\tilde V_{V-}$ 
can become negative near the horizon if $\lambda$ is sufficiently 
close to $\lambda_{c-}$, provided that the spectrum of $k_V^2$ 
extends to $k_V^2<n-1$.

\begin{figure}[t]
\centerline{\includegraphics[width=5cm]{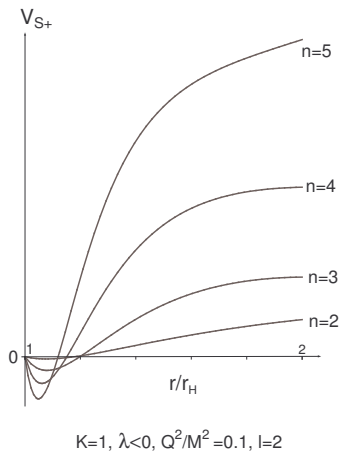}}
\caption{Examples of $V_{S+}$ for $K=1$ and $\lambda<0$.}
\label{fig:VSplus}
\end{figure}

\subsection{Scalar perturbation}

By applying the $S$-deformation to $V_{S+}$ with
\Eq{
S=\frac{f}{h_+}\frac{dh_+}{dr};\ h_+=r^{n/2-1}H_+,
}
we obtain 
\Eq{
\tilde V_{S+}=\frac{k^2 f }{2r^2 H_+}
  \left[ (n-2)(n+1)\delta x + 2\right].
}
Since this is positive definite, the electromagnetic mode $\Phi_+$ 
is always stable for any values of $K$, $M$, $Q$ and $\lambda$, 
provided that the spacetime contains a regular black hole, although 
$V_{S+}$ has a negative region near the horizon when $\lambda<0$ and 
$Q^2/M^2$ is small (see Fig. \ref{fig:VSplus}).

Using a similar transformation, we can also prove the stability of 
the gravitational mode $\Phi_-$ for some special cases. For example, 
the $S$-deformation of $V_{S-}$ with
\Eq{
S=\frac{f}{h_-}\frac{dh_-}{dr};\ h_-=r^{n/2-1}H_-
}
leads to
\Eq{
\tilde V_{S-}=\frac{k^2f}{2r^2H_-}
  \left[ 2m-(n+1)(n-2)(1+m\delta)x \right].
}
For $n=2$, this is positive definite for $m>0$. When $K=1$, 
$\lambda\ge0$ and $n=3$ or when $\lambda\ge0,Q=0$ and the horizon is 
$S^4$, from $m\ge n+2$ ($l\ge2$) and the behaviour of $x_h$ (see 
Fig. \ref{fig:xh}), we can show that $\tilde V_{S-}>0$. Hence, in 
these special cases, the black hole is stable with respect to any 
type of perturbation. 

\begin{figure}[t]
\begin{minipage}{\halftext}
\centerline{\includegraphics[width=6cm]{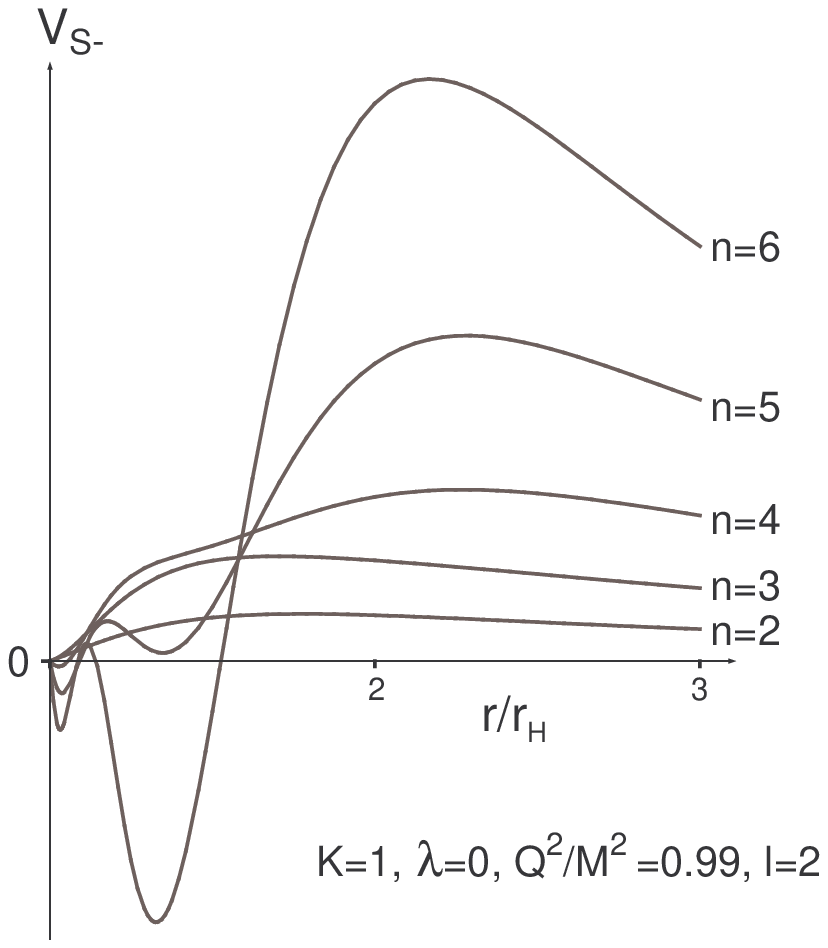}}
\end{minipage}
\begin{minipage}{\halftext}
\centerline{\includegraphics[width=6cm]{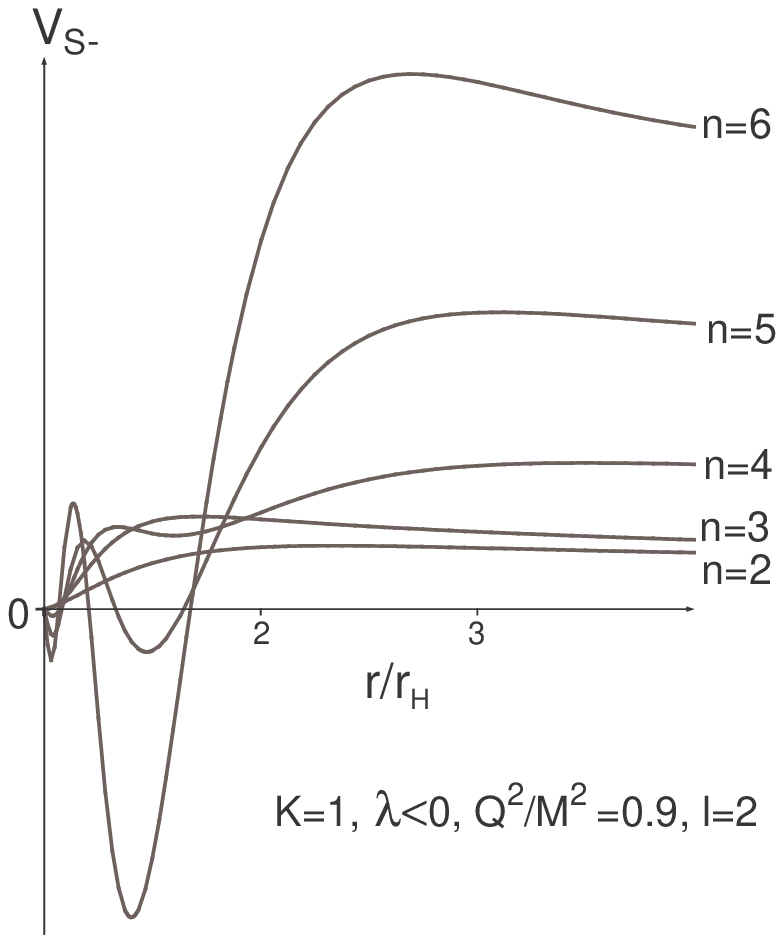}}
\end{minipage}
\caption{Examples of $V_{S-}$.}\label{fig:VSminus}
\end{figure}

However, for the other cases, $\tilde V_{S-}$ is not positive 
definite for generic values of the parameters. The $S$-deformation 
used to prove the stability of neutral black holes in Paper II is 
not effective either. This is because $V_{S-}$ has a negative region 
around the horizon for the extremal and near extremal cases, as 
shown in Fig. \ref{fig:VSminus}, and the $S$-deformation cannot 
remove this negative region if $S$ is a regular function at the 
horizon. Hence, determination of the stability for these generic 
cases with $n\ge3$ is left as an open problem.

\section{Summary and Discussion}

In the present paper, we have extended the formulation for 
perturbations of a generalised static black hole given in Paper I to 
an Einstein-Maxwell system in a generalised static spacetime with a 
static electric field, and we have shown that the perturbation 
equations for vector and scalar perturbations can be reduced to two 
decoupled second-order ODEs for the gravitational mode and the 
electromagnetic mode, irrespective of the value of the cosmological 
constant and the curvature of the horizon.  In particular, we have 
found that the coupling between the perturbations of the metric and 
the electromagnetic field produces significant modifications of the 
effective potentials for the gravitational mode and the 
electromagnetic mode when the black hole is charged. Our formulation 
also provides an extension of the corresponding formulation for the 
Reissner-Nordstrom black hole in four 
dimensions\cite{Moncrief.V1974,Zerilli.F1974,Chandrasekhar.S1983B} 
to asymptotically de Sitter and anti-de Sitter cases. 

\begin{table}[t]
\caption{Stabilities of generalised static black holes. In this 
table, ``$d$'' represents the spacetime dimension, $n+2$. The 
results for tensor perturbations apply only for maximally symmetric 
black holes, while those for vector and scalar perturbations are 
valid for black holes with generic Einstein horizons, except in the 
case with $K=1,Q=0,\lambda>0$ and $d=6$.}
\label{tbl:stability}
\begin{tabular}{l|l|c|c|c|c|c|c|}
\hline\hline
\multicolumn{2}{c|}{}& \multicolumn{2}{c|}{Tensor}
  & \multicolumn{2}{c|}{Vector}& \multicolumn{2}{c|}{Scalar}\\
\cline{3-8}
\multicolumn{2}{c|}{}&$Q=0$ & $Q\not=0$ &$Q=0$ & $Q\not=0$ 
&$Q=0$ & $Q\not=0$ \\
\hline
$K=1$& $\lambda=0$ & OK & OK & OK & OK 
     & OK 
     & $\begin{array}{l} 
         d=4,5\ \text{OK} \\ d\ge6\ \text{?} 
       \end{array}$
     \\
\cline{2-8}
     &$\lambda>0$ & OK & OK & OK & OK 
     & $\begin{array}{l} 
         d\le6\ \text{OK} \\ d\ge7\ \text{?} 
        \end{array}$
     & $\begin{array}{l}
         d=4,5\ \text{OK} \\ d\ge6\ \text{?} 
        \end{array}$
     \\
\cline{2-8}
     &$\lambda<0$ & OK & OK & OK & OK 
     &  $\begin{array}{l}
          d=4\ \text{OK} \\ d\ge5\ \text{?} 
         \end{array}$
     &  $\begin{array}{l}
          d=4\ \text{OK} \\ d\ge5\ \text{?} 
         \end{array}$
     \\
\hline
$K=0$ &$\lambda<0$ & OK & OK & OK & OK 
     & $\begin{array}{l}
         d=4\ \text{OK} \\ d\ge5\ \text{?} 
        \end{array}$
     & $\begin{array}{l}
         d=4\ \text{OK} \\ d\ge5\ \text{?} 
       \end{array}$ 
     \\
\hline
$K=-1$ &$\lambda<0$ & OK & ? & OK & ? 
     & $\begin{array}{l}
         d=4\ \text{OK} \\ d\ge5\ \text{?} 
        \end{array}$
     & $\begin{array}{l}
         d=4\ \text{OK} \\ d\ge5\ \text{?} 
        \end{array}$
     \\
\hline
\end{tabular}
\end{table}

With the help of this formulation and the method used in Paper II, 
we have analysed the stability of generalised static black holes 
with charge. The results are summarised in Table 
\ref{tbl:stability}. As shown there, maximally symmetric black holes 
are stable with respect to tensor and vector perturbations over 
almost the entire parameter range; the exceptional case corresponds 
to a rather exotic black hole, whose horizon is a hyperbolic space.  
In contrast, for a scalar perturbation, we were not able to prove 
even the stability of asymptotically flat black holes with charge in 
generic dimensions, due to the existence of a negative region in the 
effective potential around the horizon in the extremal and near 
extremal cases,  in contrast to the neutral case. Whether this 
negative ditch produces an unstable mode or not is uncertain. Hence, 
the stability of asymptotically flat and asymptotically de Sitter 
black holes for $d\ge 6$ and of asymptotically anti-de Sitter black 
holes for $d\ge5$ are left as open problems.  In connection to this, 
it should be noted that the existence of a negative region in the 
effective potential near the horizon may have a significant 
influence on the frequencies of the quasi-normal modes and the 
greybody factor for the Hawking process,  even if these black holes 
are found to be stable. 

In the present paper, we have also given explicit expressions for 
the source terms in the master equations. As mentioned in the 
introduction, this information will be necessary in the estimation 
of gravitational and electromagnetic emission from black holes in 
higher dimensions. In addition to this practical application, we 
also expect that these master equations with source terms can be 
used in the analysis of static singular perturbations of black holes 
associated with some singular source such as a string or a membrane. 
For example, if one treats the C-metric as a perturbation of a 
spherically symmetric solution, it is found that this perturbation 
obeys an equation with a string source. Hence, it is expected that 
one can obtain some information concerning higher-dimensional 
analogue of the C-metric by studying singular solutions to the 
master equation with a singular source.   

\section*{Acknowledgements}

The authors would like to thank Gary Gibbons, Sean Hartnoll and 
Toby Wiseman for conversations. 
AI is a JSPS fellow, and HK is supported by the JSPS grant No. 
15540267.

\appendix

\section{Parameter range for the existence of a regular black hole}
\label{Appendix:D}

In this section, we determine the parameter range in which the 
background metric \eqref{metric:GSBH} contains a regular black hole. 
Here, by a regular black hole, we mean a degenerate Killing horizon 
or bifurcating Killing horizons at $r=r_H$ that separate(s) a 
regular region with $f(r)>0$ and a singular region containing the 
singularity at $r=0$.  

First, we consider parameter values satisfying $\lambda\ge0$ and 
$K\le0$. In this range, if $Q=0$, there exists no horizon, because 
$f$ is negative everywhere. If $Q\neq0$, from 
\Eq{
(r^{n-1}f)'=(n-1)K r^{n-2}-(n+1)\lambda r^n
  -\frac{(n-1)Q^2}{r^{n}}<0,
}
$r^{n-1}f$ is a monotonically decreasing function of $r$. Further, 
\Eq{
r^{n-1}f \tend \left\{  
   \begin{array}{ll} +\infty, & r\tend +0\\
                 \le -2M, & r\tend +\infty
   \end{array}\right. .
}
Hence, the spacetime has no region with $f>0$ that is separated from 
the singular region by Killing horizons. Therefore, the spacetime 
contains regular black holes only for $K=1$ or $\lambda<0$.

\subsection{The case $Q=0$}

In this case, since $f\tend -\infty$ as $r\tend+0$, the spacetime 
contains a regular black hole if there is a region in which $f>0$. 
From 
\Eq{
f'=-2\lambda r + \frac{2(n-1)M}{r^n},
}
it is seen that the result depends on $\lambda$.

\subsubsection{$\lambda<0$ or $K=1$ and $\lambda=0$} 

Because $f$ is monotonically increasing and becomes positive as 
$r\tend+\infty$, $f$ has a single zero at $r=r_H$, and $f>0$ for 
$r>r_H$. Thus, there is a regular black hole.

\subsubsection{$K=1$ and $\lambda>0$} Because $f$ has a single 
maximum at $r=a$, the spacetime contains a regular black hole if and 
only if $f(a)>0$. Then, since $f'(a)=0$ is equivalent to 
\Eq{
\lambda=\frac{(n-1)M}{a^{n+1}},
} 
this condition can be written as
\Eq{
a^{n-1}>(n+1)M.
}
In terms of $\lambda$, it is expressed as
\Eq{
\lambda< \frac{n-1}{n+1}\frac{1}{[(n+1)M]^{\frac{2}{n-1}}}.
}
$f$ has two zeros, at $r=r_H$ and $r=r_c$ ($r_H<r_c$), and $f>0$ for 
$r_H<r<r_c$,

\subsection{The case $Q\not=0$}

When $Q\not=0$, since $f\tend+\infty$ as $r\tend+0$, there must 
exist a point $r=a$ such that $f'(a)=0$ and $f(a)\le0$ in order for 
the spacetime to contain a regular black hole. From
\Eq{
f'=-2\lambda r + 2(n-1)\left( \frac{M}{r^{n}}
       -\frac{Q^2}{r^{2n-1}} \right),
}
$\lambda$ is expressed in terms of $a$ as
\Eq{
\lambda =g(a):=\frac{(n-1)}{a^2}\left( \frac{M}{a^{n-1}}
       -\frac{Q^2}{a^{2n-2}} \right).
}
{}From this, $f(a)$ can be written as 
\Eq{
f(a)=K-\frac{(n+1)M}{a^{n-1}}+\frac{nQ^2}{a^{2n-2}}.
}
Hence, the condition $f(a)\le0$ requires
\Eq{
D:=(n+1)^2M^2-4KnQ^2 \ge0,
}
and under this condition, $f(a)\le0$ if $a$ satisfies 
\Eq{
\max\left( 0, \frac{(n+1)M-\sqrt{D}}{2nQ^2} \right)
 \le \frac{1}{a^{n-1}} \le \frac{(n+1)M+\sqrt{D}}{2nQ^2}.
}

Note that $g(a)$ has a maximum at $a=a_m$,  where
\Eq{
a_m^{n-1}=\frac{2nQ^2}{(n+1)M},
}
and it is monotonic everywhere except at this point. Further, 
$g(+0)=-\infty$ and $g(+\infty)=0$. Hence, $\lambda<0$ for $a<a_0$, 
where
\Eq{
a_0^{n-1}=\frac{Q^2}{M},
}
and $\lambda>0$ for $a>a_0$.

\subsubsection{$K=1$ and $\lambda=0$} 

In this case, the spacetime contains a regular black hole if and 
only if $M^2\ge Q^2$, and the horizon is at 
$r^{n-1}=r_H^{n-1}=M+\sqrt{M^2-Q^2}$. 

\subsubsection{$K=1$ and $\lambda>0$} 

Because the sign of $f'(r)$ is the same as the sign of 
$g(r)-\lambda$, the condition that there is a  point $r=a$ such that 
$f'(a)=0$ is equivalent to the relation 
\Eq{
\lambda\le \lambda_{\rm{max}}:=g(a_m)
        =\frac{(n+1)(n-1)^2}{4n^2}\pfrac{n+1}{2n}^{\frac{2}{n-1}}
         \pfrac{M^2}{Q^2}^{\frac{n+1}{n-1}}
         \frac{1}{M^{\frac{2}{n-1}}}.
}
Further, from the condition $D\ge0$, we have
\Eq{
\frac{Q^2}{M^2}\le \frac{4n}{(n+1)^2}.
}
Under these conditions, $f'(a)=0$ has in general two solutions, 
$a=a_1,a_2$ $(a_0<a_1<a_m<a_2)$, and the spacetime contains a 
regular black hole if and only if $f(a_1)\le 0<f(a_2)$. 

First, from $f(a_2)>0$, we obtain
\Eq{
a_2^{n-1}>\frac{(n+1)M+\sqrt{D}}{2} =: a_{c2}^{n-1}.
}
In terms of $\lambda$, this can be expressed as
\Eq{
\lambda < \lambda_{c2}:=g(a_{c2})=2^{\frac{n+1}{n-1}}\frac{n-1}{n}
  \frac{(n-1)M+\sqrt{D}}{[(n+1)M+\sqrt{D}]^{\frac{n+1}{n-1}}}.
}
Next, because $f(a)$ becomes minimal at $a=a_m$, from 
$f(a_0)=1-M^2/Q^2$, we have $f(a_1)\le f(a_0)\le0$ if $Q^2\le M^2$. 
Also, for $M^2<Q^2\le (n+1)^2/(4n)M^2$, $f(a_1)\le0$ is equivalent 
to 
\Eq{
a_1^{n-1}\le\frac{(n+1)M-\sqrt{D}}{2} =: a_{c1}^{n-1},
}
or, in terms of $\lambda$, to 
\Eq{
\lambda \ge \lambda_{c1}:=g(a_{c1})=2^{\frac{n+1}{n-1}}\frac{n-1}{n} 
 \frac{(n-1)M-\sqrt{D}}{[(n+1)M-\sqrt{D}]^{\frac{n+1}{n-1}}}.
}

Here, if we vary $Q^2$ with $M$ fixed, we have
\Eq{
\frac{d}{d a_{c1}}(\lambda_{c2}-\lambda_{c1})=\frac{(n-1)\sqrt{D}}{n}
  \left( \frac{1}{a_{c2}^{n+2}}- \frac{1}{a_{c1}^{n+2}}\right)<0.
}
Hence, $\lambda_{c2}-\lambda_{c1}$ is a monotonically increasing 
function of $Q$ for fixed $M$, and $\lambda_{c1}=\lambda_{c2}$ at 
$Q^2=(n+1)^2M^2/(4n)$. From these results, it follows that 
$\lambda_{c1}<\lambda_{c2}$ for $Q^2<(n+1)^2/(4n)M^2$. Further, 
$\lambda_{c1}$ vanishes at $Q^2=M^2$. Therefore, for $\lambda>0$ and 
$K=1$, the spacetime contains a regular black hole if and only if 
\Eq{
\max\left( 0,\lambda_{c1} \right) \le \lambda < \lambda_{c2}:\ 
Q^2<\frac{(n+1)^2M^2}{4n}.
}
%

\begin{table}[t]
\caption{The parameter range in which the spacetime contains a 
regular black hole.}
\label{tbl:BHcondition}
\begin{center}
\begin{tabular}{|c||c|c|c|}
\hline
  & $K=1$ & $K=0$ & $K=-1$ \\
\hline\hline
$\lambda=0$ & $Q^2\le M^2$ & $\not\exists$ & $\not\exists$ \\
\hline
$\lambda>0$ & For $Q^2\le M^2$ & & \\
            & $\lambda<\lambda_{c2}$ & $\not\exists$ & $\not\exists$ \\
            & For $M^2<Q^2<(n+1)^2M^2/(4n)$ && \\
            & $\lambda_{c1}\le\lambda<\lambda_{c2}$ && \\
\hline
$\lambda<0$ & $Q^2<M^2$ and $\lambda_{c1}\le \lambda$ 
            & $\lambda_{c0}\le \lambda$ 
            & $\lambda_{c-} \le \lambda$\\
\hline
\end{tabular}
\end{center}
\end{table}

\subsubsection{$\lambda<0$} 

In this case, $f'(a)=0$ always has a single solution, and 
$f\tend+\infty$ as $r\tend +0, +\infty$. Hence, the spacetime 
contains a regular black hole if and only if 
$f(a)\le0$. Because $a<a_0<a_m$, this condition is equivalent to  
\Eq{
\frac{1}{a^{n-1}}\le 
\frac{1}{a_c^{n-1}}\equiv\frac{(n+1)M+\sqrt{D}}{2nQ^2}, 
}
or in terms of $\lambda$,
\Eq{
\lambda \ge \lambda_{c}=\frac{n-1}{na_c^2}
    \left( K-\frac{M}{a_c^{n-1}} \right).
}
This, together with $D\ge0$, leads to the following conditions:
\Eqrsub{
&K=0: 
& \lambda>\lambda_{c0}:=-\frac{n^2-1}{n^2}\pfrac{n+1}{n}^{\frac{2}{n-1}}
        \pfrac{M^2}{Q^2}^{\frac{n+1}{n-1}}\frac{1}{M^\frac{2}{n-1}},\\
&K=-1:
& \lambda>\lambda_{c-}:=-2^{\frac{2}{n-1}}\frac{n-1}{n}
         \frac{\sqrt{D}-(n-1)M}{[\sqrt{D}-(n+1)M]^{\frac{n+1}{n-1}}},\\
&K=1:
& \lambda>\lambda_{c1}:=-2^{\frac{2}{n-1}}\frac{n-1}{n}
         \frac{\sqrt{D}-(n-1)M}{[(n+1)M-\sqrt{D}]^{\frac{n+1}{n-1}}}.
}

Finally, we determine the range of $x_H=2M/r_H^{n-1}$. In general, 
$\lambda$ is expressed in terms of $r_H$ as 
\Eq{
\lambda =\frac{1}{r_H^2}
   \left( K-\frac{2M}{r_H^{n-1}}+\frac{Q^2}{r_H^{2n-2}} \right).
}
{}From this, we have
\Eq{
\frac{d\lambda}{dr_H}=-\frac{2}{r^3}\left( 
    K-\frac{(n+1)M}{r_H^{n-1}}+\frac{n^2Q^2}{r_H^{2n-2}} \right).
}%
First, for $K=1$, this can be written in terms of $a_{c1}$ and 
$a_{c2}$ as
\Eq{
\frac{d\lambda}{dr_H}=\frac{2nQ^2}{r_H^2}
 \left( \frac{1}{a_{c1}^{n-1}}-\frac{1}{r_H^{n-1}} \right)
 \left( \frac{1}{r_H^{n-1}}-\frac{1}{a_{c1}^{n-1}} \right).
}
{}From this and the relation $a_{c2}<a_2<r_H <a_1<a_{c1}$, it 
follows that $\lambda$ is a monotonically increasing function of 
$r_H$ for fixed $M$ and $Q$. Hence, from the constraint 
$\lambda_{c1}\le \lambda <\lambda_{c2}$, we obtain
\Eq{
x_{H,\min}< x_H \le x_{H,\max},
\label{x_H:range:K=1}}
where $x_{H,\min}$ and $x_{H,\max}$ are the values of $x_H$ for 
$\lambda=\lambda_{c2}$ and $\lambda=\lambda_{c1}$, respectively:
\Eqrsub{
&& x_{H,\min}=\frac{4M}{(n+1)M+\sqrt{D}},
\label{xHmin}\\
&& x_{H,\max}=\frac{(n+1)M^2+M\sqrt{D}}{nQ^2}.
\label{xHmax}}
Next, for $K=0$ or $K=-1$, $d\lambda/dr_H$ can be written in terms 
of $a_c$ as
\Eq{
\frac{d\lambda}{dr_H}=\frac{2nQ^2}{r_H^2}
 \left( \frac{1}{a_{c}^{n-1}}-\frac{1}{r_H^{n-1}} \right)
 \left( \frac{1}{r_H^{n-1}}+\frac{\sqrt{D}-(n+1)M}{2nQ^2} \right).
}
{}From $r_H\le a\le a_c$, this is non-negative. Hence, from 
$\lambda_{c}\le\lambda<0$, we obtain
\Eq{
0 < x_H \le x_{H,\max}.
\label{x_H:range:K=0,-1}}
The allowed ranges of $x_H$ given in \eqref{x_H:range:K=1} and 
\eqref{x_H:range:K=0,-1} in the $(x_H,Q^2/M^2)$ plane are displayed 
in Fig. \ref{fig:xh}.

\section{Expressions for $\hat E^t_t$ and $\hat E_L$}
\label{Appendex:A}

We have the following expressions for $\hat E^t_t$ and $\hat E_L$:
\Eqr{
& 2\hat E^t_t= & -f \partial_r^2(X+S_T)
    +\left( \frac{n-4}{r}f-\frac{f'}{2} \right)\partial_r(X+S_T)
   \notag\\
&& +\left[\frac{k^2+(n-2)K}{r^2} +\frac{(n-2)(n-3)M}{r^{n+1}} \right.
   \notag\\
&& \quad\qquad\left.-3(n-1)\lambda +\frac{(2n-3)Q^2}{r^{2n}}\right] (X+S_T)
   \notag\\
&& -f\partial_r^2 Y 
   -\left( \frac{4f}{r}+\frac{f'}{2} \right)\partial_r Y 
   \notag\\
&& +\left[-\frac{2K}{r^2}-\frac{(n-2)(n+3)M}{r^{n+1}}
    +3\lambda +\frac{(n+1)(2n-3)Q^2}{r^{2n}} \right] Y,
}
\Eqr{
& 2\hat E_L =
  & \frac{1}{f}\partial_t^2 X+\frac{f'}{2}\partial_r X
  \notag\\
&&  +\left[ -\frac{(n-1)(n-2)M}{r^{n+1}}
      -3\lambda +\frac{n(n-1)Q^2}{r^{2n}} \right] X
   \notag\\
&& -f\partial_r^2Y 
   -\left( \frac{2f}{r}+\frac{3f'}{2} \right)\partial_r Y
   \notag\\
&&   +\left[\frac{(n-1)(n-2)M}{r^{n+1}}
    +3\lambda -\frac{3(n-1)(n-2)Q^2}{r^{2n}} \right] Y
   \notag\\
&& +\frac{2}{f}\partial_t\partial_r Z
   +\left( \frac{2}{rf}-\frac{f'}{rf^2} \right)\partial_t Z
   \notag\\
&& -f\partial_r^2 S_T 
   + \left( \frac{n-3}{r}f-f' \right)\partial_r S_T
   \notag\\
&& +\left[ \frac{(n-1)k^2}{nr^2}+\frac{2(n-1)(n-2)M}{r^{n+1}}
    \right.\notag\\
&&\quad\qquad \left.
     -3(n-1)\lambda-\frac{2(n-1)(n-3)Q^2}{r^{2n}} \right]S_T.
}
%

\section{Expressions for coefficient functions}
\label{Appendix:B}

We have the following:
\Eqr{
& P_{S1}=& \left[ -4n^4z+2n^2(n+1)x-4n(n-2)m \right]y \notag\\
&& +\left\{ 2n^2(n-1)x+4n(n-2)m+4n^3(n-2)K \right\}z
 \notag\\
&& -n^2(n^2-1)x^2+\left\{ -4n(n-2)m+2n^2(n+1)K \right\}x
\notag\\
&& +4m^2+4n^2mK, \\
& P_{S2}=& \left[ 6n^4z-n^2(n+1)(n+2)x+2n(n-4)m \right]y
\notag\\
&& -2n^4 z^2+\left\{ n^2(3n^2-n+2)x-4n(n-2)m 
   -6n^3(n-1)K \right\}z \notag\\
&&   -n^2(n+1)x^2
   +\left\{ n(3n-7)m+n^2(n^2-1)K \right\}x
  \notag\\
&& -2m^2-2n(n-1)mK,\\
& P_{S3}=& -2n^2(3n-2)z+n^2(n+1)x-2(n-2)m.
}
\Eqr{
&P_{X0} =& \left[ 4n^2(n^2-1)(n^2z+m)x
       -8n^4(n-1)(3n-2)z^2 \right.\notag\\
&& \quad \left.-32n^2(n-1)^2mz -8(n-1)(n-2)m^2 \right]y 
  \notag\\
&& +n^3(n+1)^3x^3+2n(n+1)\left\{-n^2(3n-1)(n+2)z \right.\notag\\
&&\qquad \left.    +2(n^2+n+2)m-n(n-2)(n+1)K \right\}x^2 \notag\\
&& +\left[ 4n^4(13n-7)z^2
    -4n^2\left\{ (9n^2-8n+15)m+n(n+1)(n^2-7n+8)K \right\}z\right.
    \notag\\
&& \quad\left. -4n(n-11)m^2-4n^2(n+1)(n-3)m K \right]x \notag\\
&& -8n^4(3n-2)z^3+8n^2\left\{ 4(n^2-n+1)m+ n^2(3n-4)(n-2)K\right\}z^2
  \notag\\
&& +\left\{ (16n^2-72n+16)m^2+16n^3(2n-3)mK \right\}z \notag\\
&& +8n^2m^2K +16m^3,\\
&P_{X1} =& \left[ 4n^2(n-1)z+4(n-1)m \right]y \notag\\
&& +n(n+1)^2x^2+\left\{ -8n^2z+2(3n-1)m-2n(n+1)K \right\}x\notag\\
&& +4n^2z^2-\left\{ 4(2n-1)m+4n^2(n-2)K \right\}z
   -4nmK,\\
&P_{Y0}=& \left[ 2n^4(n+1)^2x^2
      -4n^2(n+1)\left\{ n^2(5n-3)z+(n-3)m \right\}x \right.\notag\\
&&\quad\left. +8n^4(2n-1)(3n-2)z^2+16n^2(n^2-4n+2)mz
      -8(n-2)m^2 \right]y \notag\\
&& +n^3(n-1)(n+1)^2x^3 \notag\\
&&  +2n(n^2-1) \left\{ -2n^2(3n-1)z 
     +4m-n(n-2)(n+1)K\right\}x^2 \notag\\
&& +\left[ 20n^4(2n-1)(n-1)z^2
    +4n^2(n-1)\left\{(n-13)m+n(5n-8)(n+1)K \right\}z \right.\notag\\
&& \qquad \left.+12n(n-1)m^2+4n^2(n^2-1)mK \right]x \notag\\
&& -8n^4(3n-2)(n-1)z^3+16n^2(n-1)\left\{ 2m-n^2(3n-4)K \right\}z^2
  \notag\\
&& -8(n-1)m\left\{(5n-2)m+2n^3K  \right\}z,\\
&P_{Y1}=& \left[ 2n^2(n+1)x-4n^2(2n-1)z+4m \right]y \notag\\
&& +n(n^2-1)x^2-2(n-1)\left\{3n^2z+m+n(n+1)K  \right\}x
\notag\\
&& +4n^2(n-1)z^2+4(n-1)(m+2n^2K)z,\\
&P_Z=& \left[ -n^2(n+1)x+2n^2(3n-2)z+2(n-2)m \right]y \notag\\
&& +n(n+1)x^2+\left\{ n^2(3n-7)z+(4n-2)m+n(n+1)(n-2)K \right\}x
\notag\\
&& -2n^2(n-2)z^2-\left\{ (6n-4)m+2n^2(3n-4)K \right\}z 
    -2nmK.
}
Note that these functions satisfy the following relations:
\Eqrsub{
&& P_{X0}+P_{Y0}+4nH P_Z=16H^3,\\
&& P_{X1} +P_{Y1}=-4nrf H.
}
\Eqr{
&P_{XA}=& \left[ 4n^3(n-1)z+4n(n-1)m \right]y \notag\\
&& +n^2(n+1)^2x^2+4n\left\{ -2n^2z+(n-1)m-n(n+1)K \right\}x
\notag\\
&& +4n^3z^2-4n\left\{(n-1)m+n^2(n-3)K  \right\}z \notag\\
&& -4m^2-4n(n+1)mK,
}
\Eqr{
&P_{X2}=& \left[ 4(n-1)n^2z+4(n-1)m \right]y \notag\\
&& +n(n+1)^2x^2+\left\{ -8n^2z+(6n-2)m-2n(n+1)K \right\}x
\notag\\
&& +4n^2z^2-\left\{ 4(2n-1)m+4n^2(n-2)K \right\}z
   -4nmK,
}
\Eqr{
&P_{X3}=&(n+1)x+2n(n-2)z+2m ,
}
\Eqr{
&P_{YA}=&\left[ 2n^3(n+1)x-4n^3(2n-1)z+4nm \right]y \notag\\
&& +n^2(n^2-1)x^2+\left\{ -6n^3(n-1)z+4nm-2n^2(n+1)(n-2)K \right\}x
\notag\\
&& +4n^3(n-1)z^2+4n\left\{-m+n^2(2n-3)K  \right\}z
   +4nmK+4m^2,
}
\Eqr{
&P_{Y2}=&\left[ 2n^2(n+1)x-4n^2(2n-1)z+4m \right]y \notag\\
&& +n(n^2-1)x^2-2(n-1)\left\{3n^2z+m+n(n+1)K  \right\}x
\notag\\
&& +4n^2(n-1)z^2+4(n-1)( m+2n^2K)z,
}
\Eqr{
&P_{Y3}=(n-1)\left[(n+1)x-4nz \right].
}
The following relations hold for these functions:
\Eq{
P_{XA}+P_{YA}=-4n^2fH,\quad
P_{X2}+P_{Y2}=-4nfH,\quad 
P_{X3}+P_{Y3}=2H.
}

\section{Gauge-invariant treatment of the exceptional modes}
\label{Appendix:C}

In the present paper, we imposed the Einstein equation 
$F^a_a+2(n-2)F=0$ as the gauge condition for the exceptional modes 
of the scalar perturbation with $k^2=n$ for $K=1$. As discussed in 
Paper I, this gauge condition does not fix the gauge freedom 
completely, and it leaves the residual gauge freedom represented by 
$\bar\delta z^i=L\SHB^i$ satisfying 
\Eq{
\frac{1}{r^n}D\cdot(r^nDL)+\frac{n-2}{r^2}L=0.
} 
Hence, the master variables introduced in \S 
\ref{sec:ScalarPerturbation} contain unphysical degrees of freedom 
for the exceptional modes, although the master equations themselves 
are gauge-invariant. In this appendix, we express the master 
equations in terms of genuinely gauge-invariant variables.

We define all quantities for an exceptional mode corresponding to 
the gauge-invariant variables for a generic mode introduced in the 
text by the same expressions with $H_T=0$. Then, for the general 
gauge transformation 
\eqref{GaugeTrf:coord}, $X_a$ transforms as $X_a \tend X_a 
+\bar\delta X_a$ with 
\Eq{
\bar \delta X_a=-r^2D_a (L/k) +T_a.
}
{}From this, the following transformation laws of $F$ and $F^a_b$ are 
obtained:
\Eqrsub{
&F \tend F+\bar\delta F:& \bar\delta F=-rDr\cdot D(L/k)- L/k,\\
&F_{ab} \tend F_{ab}+\bar\delta F_{ab}:
  & \bar\delta F_{ab}=-D_a\left(r^2D_b \frac{L}{k}  \right)
       -D_b\left(r^2D_a \frac{L}{k}  \right).
}
In particular, we have
\Eq{
\bar\delta(F^c_c-2nF)=-2\left[ 
             r^n D\cdot\pfrac{D(L/k)}{r^{n-2}}-n\frac{L}{k} \right].
}
Hence, from \eqref{GaugeTrf:EMF} and 
\eqref{GaugeInvVar:scalar:EMF}, $\E$ and $\E_a$ transform as
\Eqrsub{
&& \bar\delta\E=-D\cdot\left(r^2E_0D(L/k)\right)
               =-qD\cdot\left( \frac{D(L/k)}{r^{n-2}} \right),
   \\
&& \bar\delta\E_a=rE_0D_a L=\frac{q}{r^{n-1}}D_aL.            
}
{}From this and the definition of $\A$, 
\eqref{BasicEq:EM:Scalar:EbyA}, we find that $\A$ transforms as
\Eq{
\bar\delta \A=q \frac{L}{k}.
}

Like these variables, the matter variables $\Sigma_{ab}, \Sigma_a$ 
and $\Sigma_L$ are gauge dependent. However, $S_a, S^a_b$ and $S_L$ 
are gauge-invariant, because they represent perturbations of 
quantities whose background values vanish, like $J_a$.

In order to proceed further, we have to treat the black-hole-type 
background case and the Nariai-type background case separately.

\subsection{Black hole background}

In the black hole background \eqref{metric:GSBH}, the gauge 
transformation of $F^a_b$ is written
\Eqrsub{
&& \bar\delta F^t_t=-\frac{2\omega^2r^2}{f}\frac{L}{k}
                   -r^2f'\frac{L'}{k},\\
&& \bar\delta F^r_t=i\omega\left[ 2r^2f\frac{L'}{k}
         +2r\left( f-\frac{rf'}{2} \right)\frac{L}{k} \right],\\
&& \bar\delta F^r_r=-2r^2f(L''/k)-r(4f+rf')(L'/k).
}
In particular, we have
\Eq{
\bar\delta \left( 2F+\frac{F^r_t}{i\omega r} \right)
 =-\frac{2H}{n}\frac{L}{k}.
}
Hence, the master variable $\Phi$ transforms as
\Eq{
\bar\delta \Phi=-2r^{n/2}\frac{L}{k}.
}

For $k^2=n$ and $K=1$, $\mu=M$ and  $\Phi_\pm$ are written
\Eqrsub{
&& \Phi_+=\frac{(n+1)MQ}{r^{n/2-1}}
       \left( \frac{\Phi}{r^{n/2}}+\frac{2\A}{q} \right),\\
&& \Phi_-=\left( 2(n+1)M-\frac{2nQ^2}{r^{n-1}} \right)\Phi
       -\frac{4nQ^2}{r^{n/2-1}}\frac{\A}{q}.
}
{}From this, we find that $\Phi_+$ is gauge-invariant. In contrast, 
as is seen from the relation
\Eq{
\Phi=\frac{\Phi_-}{2(n+1)M}+\frac{n\sqrt{\delta}}{(n+1)M}\Phi_+,
}
$\Phi_-$ is not gauge-invariant. Nevertheless, the master equation 
for $\Phi_-$ is gauge-invariant. This becomes evident if we rewrite 
this equation in terms of the gauge-invariant combinations
\Eqrsub{
&& \hat F^t_t:=F^t_t-\frac{1}{r^{n/2}}\left[ 
           \frac{r^2f'}{2}\Phi'+\left( \frac{\omega^2r^2}{f}
            -\frac{nrf'}{4} \right)\Phi \right],\\
&& \hat F^r_t:=F^r_t+\frac{i\omega r}{r^{n/2}}\left[ 
           rf\Phi'-\frac{(n-2)f+rf'}{2}\Phi\right],\\
&& \hat F^r_r:=F^r_r-\frac{1}{r^{n/2}}\Big[
         r^2f\Phi''-\frac{2(n-2)rf-r^2f'}{2}\Phi' \notag\\
&& \qquad\qquad
         +\frac{n(n-2)f-nrf'}{4}\Phi\Big],\\
&& \hat F:=F-\frac{1}{2r^{n/2}}\left[ rf\Phi'
       +\left( 1-\frac{nf}{2} \right)\Phi \right].
}
Then, taking account of the fact that the master equation was 
derived under the gauge condition $F^a_a+2(n-2)F=0$, we obtain
\Eq{
\hat F^a_a+2(n-2)\hat F
 =-\frac{1}{2(n+1)Mr^{n/2-2}f}
  \left[ S_{s-}+2n\sqrt{\delta}S_{s+}
  +2n\sqrt{\delta}(V_{s+}-V_{s-})\Phi_+ \right].
}
Here, for $k^2=n$ and $K=1$, we have 
\Eqr{
& V_{s+}-V_{s-} =& \frac{(n-1)f}{2r^2H_+^2}\big[
          \left\{ 4-2n^2(n+1)\delta x \right\}y \notag\\
&& +\delta(n^2-1)x^2+\{2n(n+1)(n-2)\delta -2n+2\}x].
\label{Scalar:Ex:Phi-}
}
Since this equation is written only in terms of gauge-invariant 
variables, it is valid in any gauge. Thus, the master equation for 
$\Phi_-$ gives an algebraic relation among the gauge-invariant 
variables $\Phi_+$, $\hat F$ and $\hat F^a_b$. 

The definition of $\Phi$ yields another relation,
\Eq{
2i\omega r \hat F+ \hat F^r_t=0.
\label{Scalar:Ex:PhiDef}
}
Further, the expressions \eqref{XYZbyPhi:RNBH} for $X$ and $Z$ 
provide two more relations:

\Eqrsubl{Scalar:Ex:XZ}{
&&r^{n-2}(\hat F^t_t-2\hat F) 
  =\frac{n^2(n-1)Q}{2(n+1)Mr^{n/2+1}}\left( 
       \frac{nP_{X+}}{2H^2} \Phi_+ -\frac{2rf}{H}\Phi_+'\right)
       +X_{s}|_{\A=0},\\
&& r^{n-2}\hat F^r_t 
 =-i\omega \frac{n^2(n-1)Qf}{(n+1)Mr^{n/2}H}\Phi_+
       +Z_{s}|_{\A=0},
}
where
\Eqr{
& P_{X+}= & \left[ (n+1)(n-2)x+2n^2z \right] y 
     +(n+1)(2n-1)x^2 \notag\\
&& -\{(3n^2+3n-2)z+(n+1)(n+2)\}x
   +2n^2z^3-2n(n-4)z.
}

The four equations \eqref{Scalar:Ex:Phi-}, \eqref{Scalar:Ex:PhiDef} 
and \eqref{Scalar:Ex:XZ} can be solved to yield expressions for 
$\hat F$ and $\hat F^a_b$ in terms of $\Phi_+$ and the 
gauge-invariant matter source terms $S_a$, $S^a_b$ and $\tilde J_a$. 
Therefore, the only dynamical gauge-invariant variable for the 
exceptional modes is $\Phi_+$. Note that we can impose the gauge 
condition $\Phi=0$, and for this gauge, $\hat F^a_b=F^a_b$ and $\hat 
F=F$ hold and $\Phi_+$ is proportional to $\A$. 

\subsection{Nariai-type background}

The case of the Nariai-type background can be treated in almost  the 
same way. First, the gauge transformations of $F^a_b$ and $F$ are 
written
\Eqrsub{
&& \bar\delta F^t_t=-2a^2\left( \frac{\omega^2}{f}\frac{L}{k} 
                   +\frac{f'}{2}\frac{L'}{k}\right),\\
&& \bar\delta F^\rho_t=2ia^2\omega\left( f\frac{L'}{k}
                -\frac{f'}{2}\frac{L}{k} \right),\\
&& \bar\delta F^\rho_\rho=-2a^2\left( f\frac{L''}{k}
               +\frac{f'}{2}\frac{L'}{k} \right),\\
&& \bar\delta F=-\frac{L}{k}.
}
For $m=0$ and $K=1$, $\sigma$ and $\mu$ have the simple expressions
\Eqrsub{
&& \sigma=\frac{n-1}{a^2} -\frac{n(n-1)Q^2}{a^{2n}},\\
&& \mu=\frac{n-1}{a^2} +\frac{n(n-1)Q^2}{a^{2n}}.
}
Hence, $\Phi_\pm$ are written
\Eqrsub{
&& \Phi_+=\frac{2(n-1)Q}{a^2}\left( F+\frac{\A}{q} \right),\\
&& \Phi_-=\frac{2(n-1)}{a^2}\left( F-\frac{nQ^2}{a^{2n-2}}
           \frac{\A}{q} \right).
}
{}From this, we find that $\Phi_+$ is gauge-invariant, while 
$\Phi_-$ is not gauge-invariant, as seen from the relation
\Eq{
F=\frac{1}{2\mu}\left( \Phi_- + \frac{nQ}{a^{2n-2}}\Phi_+ \right),
}
as in the black hole background case.

We can construct the following gauge-invariant quantities from 
$F^a_b$ and $F$:
\Eqrsub{
&& \hat F^t_t=F^t_t -2a^2\left( \frac{\omega^2}{f}F
                +\frac{f'}{2}F' \right),\\
&& \hat F^\rho_t=F^\rho_t + 2i\omega a^2\left( fF'
                 -\frac{f'}{2}F \right),\\
&& \hat F^\rho_\rho=F^\rho_\rho -2a^2\left( fF''+\frac{f'}{2}F' \right).
}
Under the gauge condition $F^a_a+2(n-2)F=0$, we have
\Eq{
\hat F^a_a=-\frac{2a^2}{f}\left[ f(fF')' 
            +\left(\omega^2+\frac{n-2}{a^2}f  \right)F\right].
}
{}From this, it follows that the master equation for $\Phi_-$ can be 
expressed in terms of the gauge invariant variables as
\Eq{
\hat F^a_a=-\frac{a^2}{\mu f}\left( S_-+\frac{nQ}{a^{2n-2}}S_+ \right)
         -\frac{nQ}{a^{2n-4}}\Phi_+.
}
Further, the equations for $X$ and $Z$ can be rewritten as
\Eqrsub{
&& \hat F^t_t
   =-\frac{nQ}{a^{2n-4}}-\frac{2}{na^{n-4}}S^\rho_\rho,\\
&& \hat F^\rho_t
   =\frac{2}{na^{n-4}}S^\rho_t.
}
The corresponding expression for $\hat F^\rho_\rho$ can be obtained 
from those for $\hat F^t_t$ and $\hat F^a_a$.  Under the gauge 
condition $F=0$, $\hat F^a_b$ coincides with $F^a_b$, and $\Phi_+$ 
becomes a constant multiple of $\A$.
 

\end{document}